\documentclass[nofootinbib,preprint,aps,12pt]{revtex4}
\usepackage[reqno]{amsmath}
\usepackage{amsthm}
\usepackage{amssymb}
\usepackage{array}
\usepackage{longtable}
\usepackage{enumerate}
\usepackage{verbatim}
\usepackage{fancyhdr}
\usepackage{graphicx}
\usepackage{bm}
\usepackage[pdftex]{hyperref}
\hypersetup{bookmarksnumbered=true,colorlinks=true,linkcolor=blue,citecolor=magenta}

\providecommand{\R}{\ensuremath{\mathbb{R}}}

\providecommand{\al}{\alpha}

\providecommand{\ga}{\gamma}
\providecommand{\Ga}{\Gamma}
\providecommand{\om}{\omega}

\providecommand{\ep}{\epsilon}

\providecommand{\sig}{\sigma}

\renewcommand{\Re}{\text{Re\,}} 
\renewcommand{\Im}{\text{Im\,}} 
\DeclareMathOperator{\Arg}{Arg} 

\DeclareMathOperator{\sgn}{sgn}
\newcommand{\EL}[1]{\Biggl\{#1\Biggr\}} 
\newcommand{\NL}[1]{\bigl\{#1\bigr\}}   

\newcommand{\EC}[1]{\Biggl[#1\Biggr]}   
\newcommand{\GC}[1]{\biggl[#1\biggr]}   
\newcommand{\MC}[1]{\Bigl[#1\Bigr]}     
\newcommand{\NC}[1]{\bigl[#1\bigr]}     

\newcommand{\EP}[1]{\Biggl(#1\Biggr)}   
\newcommand{\GP}[1]{\biggl(#1\biggr)}   
\newcommand{\NP}[1]{\bigl(#1\bigr)}     

\newcommand{\NB}[1]{\bigl|#1\bigr|}     
\newcommand{\GB}[1]{\biggl|#1\biggr|}   
\newcommand{\EB}[1]{\Biggl|#1\Biggr|}   

\newcommand{\sumf}[4]{\sum_{#1=#2}^{#3}{#4}}

\newcommand{\sumseminf}[2]{\sum_{#1=0}^{\infty}{#2}}

\newcommand{\intseminf}[2]{\int_{0}^{\,\infty}{#1\,d#2}}

\newcommand{\ket}[1]{\bigl|#1\bigr\rangle}
\newcommand{\bra}[1]{\bigl\langle#1\bigr|}
\newcommand{\ev}[1]{\bigl\langle#1\bigr\rangle}
\newcommand{\braket}[3]{\bigl\langle#1\bigr|#2\bigl|#3\bigr\rangle}
\newcommand{\Braket}[2]{\bigl\langle#1|#2\bigr\rangle}
\allowdisplaybreaks

\begin{document}
\setcounter{page}{1}
\title
{Formal Aspects of Quantum Decay}
\author
{D. F. Ram\'irez Jim\'enez}
\email{df.ramirezj@uniandes.edu.co}
\author
{N. G. Kelkar}
\email{nkelkar@uniandes.edu.co}
\affiliation{ Departamento de F\'isica, Universidad de los Andes,
Cra.1E No.18A-10, Bogot\'a, D.C., Colombia}

\begin{abstract}
The Fock-Krylov formalism for the calculation of survival probabilities 
of unstable states is 
revisited paying particular attention to the mathematical constraints on 
the density of states, the Fourier transform of which gives the survival 
amplitude. We show that it is not possible to 
construct a density of states corresponding to a purely
exponential survival amplitude.  he survival probability $P(t)$ and the autocorrelation function of the 
density of states are shown to form a pair of cosine Fourier transforms. 
This result is a particular case of the Wiener Khinchin theorem and forces 
$P(t)$ to be an even function of time which in turn forces the density of 
states to contain a form factor which vanishes at large energies. 
Subtle features of the transition regions from the non-exponential to the 
exponential at small times and the exponential to the power law decay at 
large times are discussed by expressing $P(t)$ as 
a function of the number of oscillations, $n$, performed by it. The 
transition at short times is shown to occur when the survival probability 
has completed one oscillation. The number of oscillations depend on the 
properties of the resonant state and a complete description of the evolution 
of the unstable state is provided by determining the limits on the number 
of oscillations in each region.  

\end{abstract}
\maketitle

\section{Introduction}
Spontaneous decay is an inherently quantum mechanical process. The 
likelihood for the existence of a decaying (unstable) state at a given 
point of time 
is known as the survival probability and must be calculated within the 
framework of quantum mechanics. The most intriguing fact that followed from 
the theoretical studies is that the survival probability cannot  
at all times follow the exponential decay law \cite{Khalfin} 
mostly observed in the laboratory. 
The decay law is expected to be quadratic at small times 
\cite{sudarshan,levitan,ghirardi1979,boyanov} and a power law 
at large times \cite{Fonda1,ournonexpo1, ournonexpo2, TorronMuga}. 
The theoretical claims led experimental 
nuclear and particle physicists to perform experiments (see
\cite{expoexpt} and references therein) with nuclei such as 
$^{222}$Rn, $^{60}$Co and $^{56}$Mn with half-lives ranging from hours to days. 
In spite of performing observations for several half-lives 
only an exponential decay law was measured at all times. The unique experiment 
where the non-exponential behaviour at large times was confirmed 
involved the measurement of the
luminescence decays of many dissolved organic materials after pulsed laser excitation \cite{largetexpt}.      
Experimental evidence for short time non-exponential decay was found in a
quantum tunneling experiment \cite{shorttexpt} 
where ultra-cold sodium atoms were trapped in 
an accelerating periodic optical potential created by a standing wave of 
light.
On the theoretical side, the decay law has been investigated using various 
different formalisms in literature 
(see \cite{ourpapercriticaltimes} for a 
comparison of approaches).   
Of great interest is the calculation of the critical time for 
the transition from the quadratic to the exponential at small times and 
the exponential to the power law at large times. Predictions of the critical 
times are useful in deciding the feasibility of experimentally observing a
non-exponential decay \cite{ourpapercriticaltimes}.  

One of the most commonly used formalism for the calculation of survival 
probabilities of unstable states is the method introduced by Fock and Krylov 
(FK) \cite{Fock1}. In this method, the survival amplitude (modulus-squared of 
which gives the survival probability) is evaluated as a Fourier transform of the density of states (DOS) in the resonance. Thus, the DOS is indeed 
the crucial quantity required in the calculation and must satisfy certain 
conditions \cite{Fonda1} for the correct physical behaviour of the 
survival probability. The DOS can in principle be constructed 
using the poles and residues of the resonances, 
in a model independent way as was shown in \cite{ourpaperSmatrix}. 
An essential feature of the DOS is the existence of a threshold factor 
which ensures the correct power law at large times. In \cite{ourpapercriticaltimes}, the 
present authors obtained the expressions for the DOS using formalisms other 
than the FK. 
In the present work, we revisit the calculation of the survival 
probability with the FK framework to discover some subtle features of the 
survival probability and constraints on the density of states. 
One of the 
main observations is the fact that the survival probability can only be an 
even function of time. 
The result has consequences for the standard determination of the critical 
transition time at small times by expanding an exponential in all powers 
of $t$ \cite{sudarshan,levitan}. In fact, the result we obtain is a particular 
case of the Wiener Khinchin theorem which tells us that the survival 
probability and the autocorrelation function (constructed from the density 
of states) are cosine Fourier transforms of each other. Following this result, 
we investigate the behaviour of the survival probability at small times with 
realistic examples of resonances from nuclear physics. Constructing a 
functional form of the DOS with the desired physical features we notice that 
the absence of an energy dependent form factor (which is usually 
included in a model dependent way) in the DOS can lead to unphysical 
results. Finally, the theoretical results obtained are demonstrated 
in a more visual way by applying the expressions obtained to realistic 
resonances. 

We shall begin by very briefly introducing the Fock-Krylov (FK) method 
which can in principle be used to describe the decay of any resonant state, 
whose density of states as a function of energy is known. The formalism 
has been a common tool for investigating the behaviour of survival 
probabilities in literature. This includes the 1958 paper of Khalfin
\cite{Khalfin}, \cite{sudarshan} on the Zeno effect, \cite{levitan} on 
the short time behaviour of $P(t)$, a review \cite{Fonda1} and more 
recently Ref. \cite{urbanowski} where the authors discuss the ``true face" of 
quantum decay. There exist approaches in literature which do not use the 
FK method but rely on a potential based formalism which is useful in studying 
tunneling decays (see \cite{GC2021} and references therein) and some 
interesting aspects of the latter \cite{giacosa,koide} (we refer the 
reader to \cite{anasto} for a pedagogical review).

In the next section, using the FK formalism, we show that it is not possible
to find a density of states (DOS) corresponding to a purely exponential decay. 
Section \ref{S3} derives the relation between the survival probability 
and the autocorrelation function of the DOS which leads to a particular 
case of the Wiener-Khinchin theorem. Section \ref{S4} derives the expression 
for the survival probability of a system at small times. Here we emphasize 
that a consequence of the particular form of the Wiener-Khinchin theorem 
derived earlier is the evenness of the survival probability. 
A functional form of the density of states is considered in Section 
\ref{S5} and the dependence of the results on the choice of the 
form factor appearing in the density of states is discussed in Section 
\ref{S6}. 
Expressing the survival probability $P(t)$ as a function of the number of 
oscillations performed $n$, we discover interesting features of $P(n)$ with one 
of them being that the transition time from the small time non-exponential to 
the intermediate time exponential decay law happens at a time when the 
survival probability has completed one oscillation. These results are compared 
with those obtained from other approaches in literature.   
In section \ref{Sa}, we present results for the 
critical times and behaviour of the transition regions from the 
non-exponential to the exponential at small times and the exponential 
to the power law behaviour at large times. 

\section{Density of States Associated with the Exponential Component of the Survival Probability}\label{S2}
The Fock-Krylov (FK) method has been widely used in 
literature 
for the analysis of unstable states \cite{urbanowski,mnowak,giraldi,giraldi1,Dijk,Boyanovsky}. 
We refer the reader to \cite{ourpapercriticaltimes} for details of the 
derivation and begin here with the survival amplitude given by,    
\begin{equation}\label{e1.6}
A(t)=\int_{E_{min}}^{\infty}{dE\,\rho(E)e^{-iEt}} \, ,
\end{equation}
where $\rho(E)$ is the density of states (DOS) in the resonance and is a real 
positive function. 
In the FK method, one constructs
$\rho(E)$ by rewriting the initial state in terms of the energy eigenstates of
the decay products (see Eqs (4) - (10) in \cite{ourpapercriticaltimes}). 
The commonly used Breit-Wigner distribution is an example of such a DOS. 
Though the FK method with an energy dependent DOS has been extensively used 
in literature (\cite{Khalfin,sudarshan} to quote a few), there exist other 
approaches for the evaluation of survival probabilities. We refer the 
reader to \cite{ourpapercriticaltimes} for a comparsion between different 
approaches such as the Green's function method, Jost function based formalism 
and the Fock-Krylov formalism. 
In \cite{ourpapercriticaltimes}, the survival amplitude within  
the Green's function approach was rewritten
in a form similar to that of the Fock-Krylov method to obtain, 
\[
A(t) = \int_0^{\infty}{\GC{\frac{1}{2\pi i}\,\sum_n{ C_n(k_n)\bar{C}_n(k_n)\frac{\sqrt{E}}{k_n (k_n^2 - E)}}}\, e^{-itE} \, dE},
\]
where the sum over $n$ is over all poles. 
Given the form of the above amplitude, one may identify the quantity in the
square brackets as a density of states as in Eq. (\ref{e1.6}). 
The DOS standardly used in the FK method is 
different from the quantity in the square brackets. The authors in 
\cite{ourpapercriticaltimes} showed that restricting the sum to only fourth
quadrant poles and further considering one isolated resonance, 
the above density (referred to as $\rho^{GC}$ there) is a sum of
two terms (see Eqs (69) to (72) in \cite{ourpapercriticaltimes}).
The first term is an energy derivative of the phase shift
and the second term is negligible for narrow resonances. Thus, the
Fock-Krylov and Green's function method \cite{GCPhysScripta} 
agree for isolated narrow resonances. 

\begin{figure}[htb!]
\centering
\includegraphics[scale=1.3]{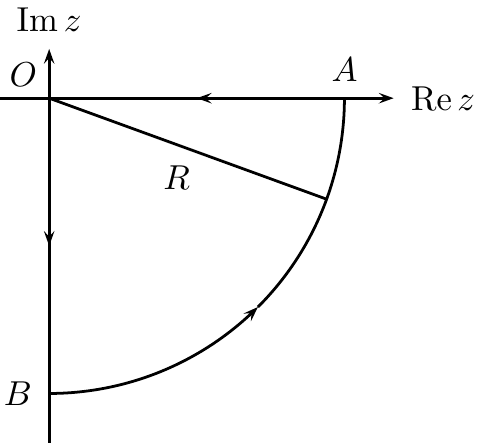}
\caption{Contour of integration for the survival amplitude}\label{fig1}
\end{figure}

The survival amplitude allows us to compute  
the so-called {\it the survival probability} $P(t)$, which measures the probability that the state of a system is in its initial state at a time, $t > 0$:
\begin{equation}\label{e1.7}
P(t)=\NB{A(t)}^2.
\end{equation}
As a natural manifestation of the normalization condition, we have, 
$A(0) = 1$ and $P(0) = 1$. 
It is known that the survival probability is split in 
three well-defined regions: the small time region where $P(t)$ follows a 
quadratic law, the intermediate time region where $P(t)$ is dominantly 
exponential and the large time region where $P(t)$ is dominated by a 
power law \cite{Fonda1,Khalfin,Nakazato1,Nakazato2}. Since some systems have long intermediate exponential regions (for instance, nuclear decays), it would be suitable to describe those decays such that the density of states gives an exponential survival probability only. However, in the following, we shall show 
that such a density of states does not exist. 
We start from the complex integral 
\begin{equation}\label{E2}
\oint_C{\rho(z)e^{-itz}\,dz},
\end{equation}
where $C$ is the contour of integration shown in Fig. \ref{fig1}. 

The exponential function (or sum of exponential ones) is obtained when the density of states has simple poles in the fourth quadrant. Since $\rho(z)$ must be a positive real function for positive real $z$, its poles have to be complex 
conjugate pairs on the right semiplane, i.e., $\rho(z)$ has simple poles at 
\begin{equation}\label{e2.0}
z=z_s\equiv\sig_s-i\om_s\sgn{(s)}/2,
\end{equation}
where $s=\pm1,\pm2,\dotsc$, $\sig_s>0$, and $\om_s>0$ for all $s$. For $s>0$ the poles are in the fourth quadrant, and for $s<0$ they are in the first 
quadrant. Let also $R(z_s)$ be the residue of $\rho(z)$ at 
the pole $z=z_s$. As a result of these conventions, we get the 
following relations:
\begin{align}
z_{s<0}&=z_{s>0}^*,\label{e2.1}\\
R(z_{s<0})&=R^*(z_{s>0}).\label{e2.2}
\end{align}
In the appendix \ref{app1}, we show from the integral \eqref{E2} that
\begin{equation*}
A(t)=-2\pi{i}\sum_{s>0}{e^{-iz_st}R(z_s)}-\int_{-i\infty}^{0}{\rho(z)e^{-izt}\,dz}.\tag{A5}
\end{equation*}
The integral \eqref{a1.5} lets us easily separate the exponential and the non-exponential component of the survival amplitude. Calling 
these respective components, $A_e(t)$ and $A_{ne}(t)$,  
from the integral \eqref{a1.5} we get:
\begin{align}
A_e(t)&=-2\pi{i}\sum_{s>0}{e^{-iz_st}R(z_s)},\label{e3.16}\\
A_{ne}(t)&=-\int_{-i\infty}^{0}{\rho(z)e^{-izt}\,dz}=-i\intseminf{\rho(-iy)e^{-ty}}{y}.\label{e3.17}
\end{align}
If the survival amplitude were purely exponential, Eq. \eqref{e3.17} would 
be zero, and therefore the density of states would vanish along the 
negative imaginary axis. Since an analytic function in a region is zero when 
the function vanishes along a continuous curve in this region \cite{Morse}, 
we deduce that $\rho(z)=0$ for all points of a domain on the $z$ plane which 
contains some segment of the negative imaginary axis. Since this domain is 
arbitrary, the density of states would be zero over the complex $z$ plane and 
then, the survival amplitude would be zero for all $t$. However, 
this is imposible because the density of states has to be different from 
zero over the real positive axis and has to have simple poles for $\Re{z}>0$. In addition to that, the survival amplitude could not be zero for all $t$ 
because the normalization condition would no longer be satisfied (apart from such an amplitude losing the meaning of a survival amplitude). 
Hence, there does not exist a density of states such that 
the survival amplitude (taken as a Fourier transform of this 
density of states) 
and the corresponding survival probability are purely exponential. 

A few words to put this result in context with the well-known seminal result
of Khalfin \cite{Khalfin} are due. Using the Paley-Wiener theorem, Khalfin 
showed that a survival amplitude given as the Fourier transform
of a semi-finite density is not attainable for arbitrary functions $M(t)$, where
the survival amplitude is given by $A(t)=M(t) \exp(\,i\phi t)$.
The result obtained in this section is in some sense a 
reverse statement of what Khalfin states. 
We show that assuming a purely exponential survival probability,
$|A(t)|^2$, it is not possible to find a density of states, the Fourier
transform of which gives the survival amplitude as in our Eq. (\ref{e1.6}).
The result is relevant since most observed decays are exponential and it
would be desirable to look for such a density of states. 
Apart from this, it has consequences for the calculation of the 
autocorrelation function which we shall see later in Section \ref{S3}. 
It is worth noting that the above result is based on a simple argument 
of analytic continuation and does not rely on the Paley-Wiener theorem.


Eq. \eqref{a1.5} lets us descompose the survival probability, $P(t)$, in a 
suitable way. Therefore, from the definition of $P(t)$  
and Eq. \eqref{a1.5}, we can write 
\begin{equation}\label{e2.3}
P(t)=|A(t)|^2=|A_e(t)|^2+|A_{ne}(t)|^2+2\,\Re{\NC{A_e^*(t)A_{ne}(t)}},
\end{equation}
If we do not focus on the non-exponential behaviour at very small times, 
we can see that the survival probability is split into three 
terms: the first one comes from the exponential component of $A(t)$, the second one comes from the non-exponential component of $A(t)$ (at large times) 
and the last one is nothing but the interference between the components 
of $A(t)$. These terms are denoted respectively as $P_e(t)$, $P_{ne}(t)$ and $P_i(t)$ and are called (in the same order) the exponential, non-exponential and interference survival probability. From the Eqs \eqref{e3.16} and \eqref{e3.17}, those terms are given by:
\begin{align}
P_e(t)&\equiv|A_e(t)|^2=4\pi^2\GB{\sum_{s>0}{e^{-iz_st}R(z_s)}}^2,\label{e3.18}\\
P_{ne}(t)&\equiv|A_{ne}(t)|^2=\EB{\int_{-i\infty}^{0}{\rho(z)e^{-izt}\,dz}}^2,\label{e3.19}\\
P_i(t)&\equiv2\,\Re{\NC{A_e^*(t)A_{ne}(t)}}=-4\pi\sum_{s>0}{\Re{\EC{iR^*(z_s)\int_{-i\infty}^{0}{\rho(z)e^{-i(z-z_s^*)t}\,dz}}}}.\label{e3.20}
\end{align}
This particular descomposition of the survival probability is useful 
for analyzing the transition of the decay law from the exponential 
to the power law at large times. 
The analysis of the latter transition using this splitting can be seen 
in \cite{ourpapercriticaltimes}. 
In section \ref{Sa} we shall study the small time transition region.
Though the above division of regions is in general valid, there exist 
exceptions where the decay law is non-exponential at all times. This happens 
in the case of broad resonances, of which the $\sigma$ meson is a good example 
\cite{ournonexpo2}. Other examples can be found in \cite{GCPRA73,GCPRA93}. 
  
\section{Survival Probability in terms of the autocorrelation of the density of states}\label{S3}
Although it is easier to compute the survival probability simply as the modulus squared of  the survival amplitude, it would be desirable to get an expression involving some function of the density of states. We shall show that this function is the autocorrelation function of the density of states. Furthermore, the expression will be a particular case of the Wiener-Khinchin theorem\footnote{For details about this theorem, see \cite{Allen} and \cite{Brandwood}.}. Using \eqref{e1.7}, 
\begin{equation}\label{e3.1}
P(t)=A(t)A^*(t)=\int_{E_{\text{min}}}^{\infty}{\int_{E_{\text{min}}}^{\infty}{\rho(E)\rho(E')e^{i(E'-E)t}\,dEdE'}}.
\end{equation}
Making the change of variables $E=x$ and $E'-E=y$, the new region of integration $S$ is given by
$S=\NL{(x,y)\in\R^2:x\geq E_{\text{min}}$ ,$x+y\geq {E_{\text{min}}}}$.
Integrating first over $x$ and then over $y$, the integral \eqref{e3.1} is split in two integrals:
\begin{equation}\label{e3.3}
P(t)=\int_{0}^{\infty}{\,dy\,e^{\,iyt}\int_{E_{\text{min}}}^{\infty}{\,dx\,\rho(x)\rho(x+y)}}+
\int_{-\infty}^{0}{\,dy\,e^{\,iyt}\int_{E_{\text{min}}-y}^{\infty}{\,dx\,\rho(x)\rho(x+y)}}.
\end{equation}
Making in the second double integral another change of variables, i.e., $y'=-y$ and $x'=x+y$; the region of integration is transformed to  
one given by $x'\geq E_{\text{min}}$ and $y'\geq0$. Thus,
\begin{equation}\label{e3.4}
P(t)=\int_{0}^{\infty}{\,dy\,e^{\,iyt}\int_{E_{\text{min}}}^{\infty}{\,dx\,\rho(x)\rho(x+y)}}+
\int_{0}^{\infty}{\,dy'\,e^{-iy't}\int_{E_{\text{min}}}^{\infty}{\,dx'\,\rho(x'+y')\rho(x')}}.
\end{equation}
Now, the integrals can be added. Therefore,
\begin{equation}\label{e3.5}
P(t)=\int_{0}^{\infty}{\,dy\,2\cos{yt}\int_{E_{\text{min}}}^{\infty}{\,dx\,\rho(x)\rho(x+y)}}.
\end{equation}
The survival probability is then the cosine Fourier transform of the function
\begin{equation}\label{e3.6}
{\cal R}(y)=2\int_{E_{\text{min}}}^{\infty}{\,dx\,\rho(x)\rho(x+y)}.
\end{equation}  
${\cal R}(y)$ is the {\it autocorrelation function} of the density of states. In order to obtain the autocorrelation function as an inverse cosine 
Fourier transform of $P(t)$, we multiply Eq. \eqref{e3.5} by $\cos{y't}$ and then we integrate with respect to $t$ between $t=0$ and $t=\infty$ so that  
\begin{align}
\intseminf{P(t)\cos{y't}}{t}
&=\int_{0}^{\infty}{\,dy\GC{\intseminf{\cos{yt}\cos{y't}}{t}}\cdot2\int_{E_{\text{min}}}^{\infty}{\,dx\,\rho(x)\rho(x+y)}}\notag\\
&=\int_{0}^{\infty}{\,dy\,\frac{\pi}{2}\MC{\delta{\NP{y-y'}}+\delta{\NP{y+y'}}}\cdot2\int_{E_{\text{min}}}^{\infty}{\,dx\,\rho(x)\rho(x+y)}}\notag\\
=&\pi\int_{E_{\text{min}}}^{\infty}{\,dx\,\rho(x)\rho(x+y)}=\frac{\pi}{2} 
{\cal R}(y).\label{e3.7}
\end{align}
To summarize, the survival probability 
and the autocorrelation function of the density of states are a pair of cosine Fourier transforms given by 
\begin{align}
P(t)&=\intseminf{{\cal R}(y)\cos{yt}}{y},\label{e3.8}\\
{\cal R}(y)&=\frac{2}{\pi}\intseminf{P(t)\cos{yt}}{t}.\label{e3.9}
\end{align}
A word of caution about the lower limit of integration of the autocorrelation 
function of the density of states is in order here. Without loss of generality, we take its value as $E_{min}=0$. This is equivalent to shifting the origin 
of the energy scale at the threshold of the system and is translated in a change of variable in the integral \eqref{e3.6} such that  
\begin{equation}\label{e3.10}
{\cal R}(y)=2\int_{0}^{\infty}{\,dx\,\rho(x)\rho(y+x)}.
\end{equation}
Eqs. \eqref{e3.8} and \eqref{e3.9} still remain valid. 

Although the aim of this work is to study the characteristic features of the 
survival probability of an unstable quantum system, some applications of 
the autocorrelation function are worth mentioning. 
In the case of a system with a discrete spectrum, it is simple to compute the autocorrelation function by taking the Fourier transform of the survival probability. Let $\NL{\ket{n}}$ be a discrete eigenstate of a system with Hamiltonian $H$ and initial state $\ket{i}$\footnote{In order to avoid any confusion related 
to the notation for the initial state used at the beginning of this article, 
this state is denoted as $\ket{i}$ only in this section.}, and let $\NL{E_n}$ 
be their corresponding eigenenergies:
$H\ket{n}=E_n\ket{n}$,
where $n=0,1,2,\dotsc$ and the discrete energies are labeled such that $E_0<E_1<E_2\cdots$. The survival amplitude and probability are then given as 
\begin{align}
A(t)&=\braket{i}{e^{-iHt}}{i}=\sum_{n=0}^{\infty}{\NB{\Braket{i}{n}}^2e^{-iE_nt}},\label{e3.12}\\
P(t)&=|A(t)|^2=\sum_{n=0}^{\infty}{\sum_{m=0}^{\infty}{{\NB{\Braket{i}{n}}^2\,\NB{\Braket{i}{m}}^2\,e^{-i(E_n-E_m)t}}}}\notag\\
&=\sum_{n=0}^{\infty}{\NB{\Braket{i}{n}}^2}
+2\sum_{n=1}^{\infty}{\sum_{m=0}^{n-1}{\NB{\Braket{i}{n}}^2\,\NB{\Braket{i}{m}}^2\,\cos{(E_n-E_m)t}}}.\label{e3.13}
\end{align}
The corresponding autocorrelation function can be computed by inspection: Eq. \eqref{e3.13} is obtained from Eq. \eqref{e3.8} by multiplying the 
first sum on the right hand side of \eqref{e3.13} by $\delta(y)$ and 
substituting $\cos{(E_n-E_m)t}$ by $\delta{\NC{y-(E_n-E_m)t}}$. Therefore,
\begin{equation}\label{e3.14}
{\cal R}(y)=\delta{(y)}\sum_{n=0}^{\infty}{\NB{\Braket{i}{n}}^2}
+2\sum_{n=1}^{\infty}{\sum_{m=0}^{n-1}{\NB{\Braket{i}{n}}^2\,\NB{\Braket{i}{m}}^2\,\delta{\NC{y-(E_n-E_m)t}}}}.
\end{equation}
Since the autocorrelation function is a convolution-like function, 
the authors in \cite{caos1,caos2,caos3} use Eq. \eqref{e3.14} together with random matrix theory for finding signatures of classical chaos in quantum systems. 
They do so by comparing numerical results with experiment. 

We pointed out in Section \ref{S2} about the impossibility of deriving a density of states corresponding to a purely exponential survival probability. 
This implies that we would not be able to use the autocorrelation 
function of the density of states given by Eq. \eqref{e3.10} if we wanted 
to identify the exponential contribution of the survival probability. 
Even if we showed that it is impossible to find a 
density of states corresponding to a purely exponential survival probability, 
one can always find a density of states 
which leads to a survival probability with an exponential component in addition to the non-exponential ones as given 
in Eqs \eqref{e3.16} and \eqref{e3.18}. 
In spite of having found a functional form for the density of states, 
the general case of the exponential component of a sum consisting of 
several poles is difficult to work with. Let us therefore consider the 
simpler case of a narrow isolated 
resonance with a long intermediate region of exponential decay. 
In this case, P(t) can be approximated roughly by its exponential component. 
Using $z_s$ as given by \eqref{e2.0}, 
$A_e(t)$ and $P_e(t)$ from \eqref{e3.16} and \eqref{e3.18} can be written as
\begin{align}
A_e(t)&=-2\pi{i}{e^{-iz_st}R(z_s)},\\
P_e(t)&=4\pi^2|R(z_s)|^2e^{-2\Im{z_s}t}=4\pi^2|R(z_s)|^2e^{-{\om_s}t},
\end{align}
and Eq. \eqref{e3.9} gives us the approximate autocorrelation function:
\begin{equation}
{\cal R}(y)=\frac{2}{\pi}\intseminf{P(t)\cos{yt}}{t}\approx 8\pi\om_s\,{\frac{|R(z_s)|^2}{y^2+\om_s^2}}.
\end{equation}
If we wish to take other resonances into account, 
we can improve the approximation of the autocorrelation function. From 
Eq. \eqref{e3.18}:
\begin{equation}\label{e3.21}
P(t)\approx P_e(t)=4\pi^2\sum_{s>0}{\sum_{s'>0}{R(z_s)R^*(z_{s'})\,e^{-i(z_s-z_{s'}^*)t}}}.
\end{equation}
Using Eq. \eqref{e3.9}, the autocorrelation function is approximately equal to:
\begin{multline}\label{e3.22}
{\cal R}(y)=\frac{2}{\pi}\Re{\intseminf{P(t)e^{-iyt}}{t}}\approx8\pi\Im{\sum_{s>0}{\sum_{s'>0}{\frac{R(z_s)R^*(z_{s'})}{y+z_s-z_{s'}^*}}}}\\
=8\pi\sum_{s>0}{|R(z_s)|^2\,\frac{\om_s}{y^2+\om_s^2}}
+16\pi\Im{\sum_{s>0}{\sum_{s'>s}{\frac{R(z_s)R^*(z_{s'})}{y+\sig_s-\sig_{s'}-i\NP{\om_s+\om_{s'}}/2}}}}.
\end{multline}
\section{Survival probability of a system for small times}\label{S4}
A consequence of Eq. \eqref{e3.8} is the evenness of $P(t)$. This 
property can be checked directly from the definition of the survival 
probability:
\begin{equation}\label{e4.1}
P(t)=A(t)A^*(t)=\braket{0}{e^{-iHt}}{0}\braket{0}{e^{-iHt}}{0}^*=\braket{0}{e^{-iHt}}{0}\braket{0}{e^{iHt}}{0}.
\end{equation}
Expanding the temporal evolution operators in a Taylor series:
\begin{align}
P(t)&=\EC{\sumseminf{n}{\frac{(-i)^n}{n!}\braket{0}{H^n}{0}t^n}}\EC{\sumseminf{m}{\frac{i^m}{m!}\braket{0}{H^m}{0}t^m}}\notag\\
&=\sumseminf{n}{\frac{(it)^n}{n!}\sumf{m}{0}{n}{(-1)^m\binom{n}{m}\braket{0}{H^m}{0}\braket{0}{H^{n-m}}{0}}}\notag\\
&=\sumseminf{n}{\frac{i^n}{n!}\,p_n(H)t^n},\label{e4.2}
\end{align}
where $p_n(H)$ is given by:
\begin{equation}\label{e4.3}
p_n(H)=\sumf{m}{0}{n}{(-1)^m\binom{n}{m}\braket{0}{H^m}{0}\braket{0}{H^{n-m}}{0}}.
\end{equation}
The sum given by \eqref{e4.3} has the form
$\sumf{m}{0}{n}{(-1)^mA_mA_{n-m}}$,
with $A_m$ given by 
\begin{equation}
A_m=\sqrt{\dbinom{n}{m}}\braket{0}{H^m}{0}.
\end{equation}
Since the above sum is zero for $n$ odd, 
$p_{n}(H)$ is zero for $n$ odd. Thus, 
\begin{equation}\label{e4.4}
P(t)=1+\sum_{q=1}^{\infty}{\frac{(-1)^q}{(2q)!}p_{2q}(H)\,t^{2q}}.
\end{equation}
Summarizing the above, we can say that not only is the survival probability 
an even function, but also its Taylor expansion around $t=0$ contains even 
powers of $t$ only. As a consequence, 
$P'(0) = 0$ and 
\begin{equation}
P''(0) = -p_2(H)=-2\ev{(\Delta H)^{2}}_0<0,\label{e4.6}
\end{equation}
where $\ev{(\Delta H)^{2}}_0=\ev{H^2}_0-\ev{H}_0^2$. Furthermore, 
Eq. \eqref{e4.4} implies that the survival probability must follow a 
quadratic law for small times, meaning, 
\begin{equation}\label{e4.6a}
P(t)=1-t^2\ev{(\Delta H)^2}_0+O(t^4).
\end{equation}
Another implication of the above result is that it is not possible  
to have a purely exponential survival probability because the Taylor 
expansion of the exponential survival probability around $t=0$ has both even and odd powers of $t$. We can see this fact if we expand the Eq. \eqref{e3.16} in Taylor series:
\begin{equation}\label{e4.6a1}
A_e(t)=-2\pi i\sum_{s>0}{R(z_s)\sum_{n=0}^{\infty}{\frac{(-iz_st)^n}{n!}}}=\sum_{n=0}^{\infty}{\frac{(-it)^n}{n!}\sum_{s>0}{(-2\pi i)z_s^nR(z_s)}}=\sum_{n=0}^{\infty}{\frac{(-it)^n}{n!}\,B_n},
\end{equation}
where $B_n=-2\pi i\sum_{s>0}{z_s^nR(z_s)}$. Therefore, the corresponding 
Taylor expansion for $P_e(t)$ will be:
\begin{equation}\label{e4.6a2}
P_e(t)=\NB{A_e(t)}^2=\sum_{n=0}^{\infty}{\frac{(it)^n}{n!}\sum_{m=0}^{n}{(-1)^m}\binom{n}{m}B_mB_{n-m}^{*}}=
\NB{B_0}^2-2t\,\Im{\NP{B_0B_1^*}}-t^2\Re{\NP{B_0B_2^*-|B_1|^2}}+\cdots.
\end{equation}
Eq. \eqref{e4.6a2} lets us interpret the role of the non-exponential and 
interference terms of the survival probability. 
Adding the Taylor expansion of both functions to the expansion of $P_e(t)$, 
the normalization condition and the evenness of the survival probability 
should be ensured. We shall see that this actually happens. 
Since the Taylor expansion of the non-exponential and interference terms of 
the survival probability depend on the Taylor expansion of the 
non-exponentical survival amplitude, 
we can expand $A(t)=\braket{0}{e^{-iHt}}{0}$ and use Eq. \eqref{e4.6a1}. 
Thus, we get the required expansion, i.e., 
\begin{equation}\label{e4.6a3}
A_{ne}(t)=A(t)-A_{e}(t)=\sum_{n=0}^{\infty}{\frac{(-it)^n}{n!}\NP{\ev{H^n}_0-B_n}}.
\end{equation}
Therefore, the Taylor expansion of the non-exponential survival probability is
\begin{multline}\label{e4.6a4}
P_{ne}(t)=\NB{A_{ne}(t)}^2=\sum_{n=0}^{\infty}{\frac{(it)^n}{n!}\sum_{m=0}^{n}{(-1)^m\binom{n}{m}\NP{\ev{H^m}_0-B_m}\NP{\ev{H^{n-m}}_0-B_{n-m}^*}}}\\
=\NB{1-B_0}^2-2t\,\Im{\MC{\NP{1-B_0}\NP{\ev{H}_0-B_1^*}}}-
t^2\,\Re{\MC{\NP{1-B_0}\NP{\ev{H^2}_0-B_2^*}-\NB{\ev{H}_0-B_1}^2}}+\cdots.
\end{multline}
In addition, the Taylor expansion of the interference part of the 
survival probability is
\begin{multline}\label{e4.6a5}
P_{i}(t)=2\Re{\NC{A_{e}(t)A_{ne}^*(t)}}
=2\Re{\sum_{n=0}^{\infty}{\frac{(it)^n}{n!}\sum_{m=0}^{n}{(-1)^m\binom{n}{m}B_m\NP{\ev{H^{n-m}}_0-B_{n-m}^*}}}}\\
=2\Re{\NC{B_0(1-B_0^*)}}-2t\,\Im{\MC{B_0\NP{\ev{H}_0-B_1^*}-B_1(1-B_0^*)}}\\-
{t^2}\,\Re{\MC{B_2-2\Re{\NP{B_0B_2^*}}+B_0\ev{H^2}_0-2B_1\NP{\ev{H}_0-B_1^*}}}+\cdots.
\end{multline}
It is gratifying to find that adding the 
Eqs. \eqref{e4.6a2}, \eqref{e4.6a4} and \eqref{e4.6a5}, 
we obtain after some algebra the Taylor expansion \eqref{e4.4}. 
Note that the expansions \eqref{e4.6a4} and \eqref{e4.6a5} are given in terms 
of the poles and residues of the DOS and 
the expectation values of the integer powers of the Hamiltonian at 
the initial state. 

Finally, owing to the condition that the coefficients $p_{n}\NP{H}$ 
have to be finite, some restrictions must be taken into account when the Fock-Krylov formalism is used. The coefficients $p_{n}\NP{H}$ are finite if the expectation value of $H^n$ at 
the initial state exists for all $n=0,1,2,\dotsc$. 
\begin{equation}\label{e4.7}
\ev{H^n}_0=\braket{0}{H^n}{0}=\bra{0}H^n\int_{0}^{\infty}{dE\,\int{db\,\ket{E,b}\Braket{E,b\,}{0}}}
=\intseminf{E^n\rho(E)}{E}<\infty,\quad n=0,1,\dotsc
\end{equation}
The conditions \eqref{e4.7} claim that the density of states is such that 
these integrals must converge for each value of $n$, and hence we infer 
that the convergence of the integrals will be possible if there exists 
a real and positive function $g(E)$ such that it is part of  
the density of states. This function is indeed the {\it form factor} 
introduced often in literature. Since there are no analytic methods 
for obtaining the form factor, it is common to see 
phenomenological procedures 
in the literature (see for instance \cite{Fock1,Fonda1,ourpaperSmatrix,
ournonexpo1,ournonexpo2,Brzeski}). We note that the 
existence of the form factor is a consequence of the evenness of the 
survival probability.

A word of caution regarding the moments of $H$ is in order before ending this
section. Assuming a general
short time dependence of the form, $A \sim  1 + b t^c$, where $b$ and $c$ are
finite constants and requiring the finiteness of the moments of $H$, 
it was noted in \cite{muga} that the derivatives 
\begin{align*}
\frac{dA}{dt}\biggr|_{t=0}&=-\frac{i}{\hbar}\braket{\Psi}{H}{\Psi}=bct^{c-1}\biggr|_{t=0},\\
\frac{d^2A}{dt^2}\biggr|_{t=0}&=-\frac{i}{\hbar^2}\braket{\Psi}{H^2}{\Psi}=bc(c-1)t^{c-1}\biggr|_{t=0},
\end{align*}
rule out the possibility of the short time $t^{1/2}$ behaviour 
(sometimes found in literature such as \cite{sudarshan} and 
\cite{garciaisolated}) of 
$A(t)$ since it implies an infinite time derivative of $A$ at $t=0$. For the 
derivatives to be finite, $c \ge 1$.  
We also note that studies of the short time behaviour within a quantum 
field theoretic (QFT) 
 approach as in \cite{maiani1,maiani2} find that the energy 
uncertainty which depends on the first two moments of $H$ is infinite. 
In a more recent QFT based calculation, however, the author 
introduces a cut-off parameter $\Lambda$ such that
there is no divergence for times $t \le 1/\Lambda$. 
This time scale determines the renormalization of the bare state and
formation of the quasiparticle state. In connection with the anti-Zeno
effect, the author also determines the energy uncertainty at a given time. 
A discussion of the finiteness of the moments and the 
short time behaviour within solvable models can be 
found in \cite{cordero}, where the authors found that the expansion 
of the survival probability in terms of resonant states predicts the 
possibility of a $t^{3/2}$ short time behavior, which follows from the
fact that in general the energy moments of the Hamiltonian may diverge. 
The authors considered the expression, $P(t) = 1 - (t/\tau^*)^{\theta}$, 
with parameters, $\theta$ and $\tau^*$ to adjust the short time behaviour 
of calculations using experiment. Two sets, $\theta$ = 2, $\tau^*$ = 12.55 
$\mu s$ and $\theta$ = 3/2 and $\tau^*$ = 23.15 $\mu s$ were found to agree
with the data at small times \cite{shorttexpt}.

\section{Density of States for Continuum Spectra}
\label{S5}
All the results obtained so far in this work indicate that knowing the 
poles and residues of the density of states and its form factor as well 
is necessary if we wish to construct $\rho(E)$ through 
the Mittag-Leffler theorem. The latter affirms that a meromorphic function 
can be constructed by knowing its poles and residues. Since the calculation 
of the survival amplitude involves a Fourier transform which must be 
performed by going over to the complex energy plane, we must consider the 
behaviour of the density of states in the complex energy plane.  
Theoretical evidence suggests that the large time behavior of 
the survival probability follows a power law that comes from a branch point 
of $\rho(E)$ at $E=0$ such that $|\Arg{E}|<\pi$. This can be established  
by supposing that the density of states has an asymptotic expansion 
around $E=0$ in the form
\begin{equation}\label{e5.1}
\rho(E)\sim E^{\nu}\sum_{n=0}^{\infty}{\beta_n E^n},\quad E\to 0^+.
 \end{equation} 
Substituting Eq. \eqref{e5.1} in Eq. \eqref{a1.5} and applying the 
Watson's lemma \cite{Ablowitz}, 
we find that the survival amplitude and 
probability have, for larges times, the asymptotic expansions
\begin{align}
A(t)&\sim\frac{1}{i}\sum_{n=0}^{\infty}{(-i)^{\nu+n}\beta_n\,\frac{\Ga\NP{\nu+n+1}}{t^{\nu+n+1}}},\label{e5.2}\\
P(t)&=|\beta_0|^2\frac{\NB{\Ga\NP{\nu+1}}^2}{t^{2\nu+2}}+O\NP{t^{-2\nu-3}}.\label{e5.3}
\end{align}
It is known from scattering theory that at large times $P(t) \propto 
t^{-(2l+3)}$ \cite{Fonda1} for a resonance in the $l^{th}$ partial wave. 
This would imply $\nu = l + 1/2$ which as we will see later is consistent with 
conditions imposed on $\nu$. 
In Section \ref{S3} we noted that the evenness of $P(t)$ 
follows from the Wiener-Khinchin theorem and indeed the small time behaviour 
was consistent with this requirement. 
However, the large time $P(t) \propto t^{-(2l+3)}$ cannot be an even function. 
Such a strange behavior is however common in many asymptotic expansions of
even functions. As a first example, consider the function 
$f(z)=(z^6+z^2+1)^{-1/2}$, which, for large values of $|z|$, behaves as 
$f(z)\sim z^{-3}$. A second example is provided by the Bessel function of 
the first kind of order $2n$, $J_{2n}(x)$ whose asymptotic expression for 
large $x$ is $\sqrt{\frac{2}{\pi\,x}}\cos{(x-\frac{\pi}{4}-n\pi)}$ 
\cite{Lebedev}. Finally, the integral
\[
g(u)=\int_{u^2}^{\infty}{\frac{e^{\,it}}{t^{1/2}}\,dt}
\]
has the asymptotic form $g(u)\sim ie^{\,iu^2}/u$ for large $u$ \cite{copson}.
It is like these functions forget how they were raised in $t=0$ and change 
completely as $t \to \infty$. 
We must note that in spite of the above, there is in general no 
contradiction with the result following from the Wiener-Khinchin theorem since 
it is the total survival probability (and not just a non-exponential part) 
which must be an even function.

Eq. \eqref{e5.3} shows that the nature of the branch point is determined by 
the exponent of the power law and this feature must be included if we want 
to obtain an expression for the density of states. In summary, the density 
of states can be built if we know 
(i) its poles and corresponding residues, 
(ii) its form factor and 
(iii) the threshold factor $\nu$ which also defines 
the exponent $2\nu+2$ of the survival probability for large times.
\\
Since the density of states can be decomposed into the product of 
$E^{\nu}$, with $\nu>0$, an analytic form factor $g(E)$ with $g(0)\neq0$, 
and a meromorphic function whose poles are the same as those of 
the density of states, the deduction of a generic expression for the 
density of states starts from the function
\begin{equation}\label{e5.4}
   F(z)=\frac{z^{-\nu}}{g(z)}\,\frac{\rho(z)}{z-E},
   \end{equation}   
which is the meromorphic component of the density of states, is 
analytic at the origin and has simple poles at $z=z_s$ and $z=E$. 
Let $C_N$ be the circle $|z|=R_N$ such that $|z_s|<R_N$ for 
$s=1,2,\dotsc N$ and contains the pole $z=E$. From the residue theorem:
\begin{equation}\label{e5.5}
\frac{1}{2\pi i}\oint_{C_N}{\frac{z^{-\nu}}{g(z)}\,\frac{\rho(z)}{z-E}\,dz}=\frac{E^{-\nu}}{g(E)}\,\rho(E)-\sum_{|s|\leq N}{\frac{1}{z_s^{\nu}g(z_s)}\,\frac{R(z_s)}{E-z_s}}.
\end{equation}
If $z^{-\nu}\rho(z)/g(z)=O\NP{|z|^{-\delta}}$ for $|z|\to\infty$ and $\delta>0$, the integral vanishes when $N$ tends to infinity. Hence,
\begin{equation}\label{e5.6}
\rho(E)=E^\nu\,g(E)\sum_{s}{\GC{-\frac{R(z_s)}{g(z_s)}}\,\frac{1}{z_s^{\nu}}\,\frac{1}{z_s-E}}.
\end{equation}
Since the density of states has to be a real function, the form factor must 
have the following property in the complex $z$-plane:
$g(z_{s<0})=g^*(z_{s>0})$.
Using this property with Eqs \eqref{e2.1} and \eqref{e2.2}, we can write an 
alternative form for $\rho(E)$:
 \begin{equation}\label{e5.7}
\rho(E)=\frac{1}{2}\,E^\nu\,g(E)\,{\sum_{s}{\frac{\ga(z_s)}{z_s^{\nu}}\,\frac{1}{z_s-E}}}=E^\nu\,g(E)\,\Re{\sum_{s>0}{\frac{\ga(z_s)}{z_s^{\nu}}\,\frac{1}{z_s-E}}},
\end{equation}
where $\gamma(z_s)$ is defined as
\begin{equation}\label{e5.7a}
\gamma(z_s)=-2\,\frac{R(z_s)}{g(z_s)},
\end{equation}
and due to Eq. \eqref{e2.1} and the property of the form factor, it satisfies 
$\gamma(z_{s<0})=\ga^*(z_{s>0})$.
Since the density of states satisfies the normalization condition, we have
\begin{equation}\label{e5.8}
\intseminf{\rho(E)}{E}=\Re{\sum_{s>0}{\frac{\ga(z_s)}{z_s^{\nu}}\,\intseminf{\frac{E^\nu}{z_s-E}\,g(E)}{E}}}=1.
\end{equation}
The density of states is such that the integrals \eqref{e4.7} 
will be finite, i.e., 
\begin{equation}\label{e5.8d}
\Re{\sum_{s>0}{\frac{\ga(z_s)}{z_s^{\nu}}\,\intseminf{\frac{E^{\nu+n}}{z_s-E}\,g(E)}{E}}}<\infty,\quad n=1,2,\cdots.
\end{equation}
Since the non-exponential survival probability given by Eq. \eqref{e3.17} is 
proportional to the Laplace transform of the density of states on the 
negative imaginary axis, by the existence theorem of the Laplace transform, 
if there exist real constants $M$ and $\al$ such that $\rho(-iy)$ has 
exponential order $\al$, i.e.,
$\NB{\rho(-iy)}\leq Me^{\,\al y}$,
the integral converges for $t> \al$ \cite{Schiff}. From Eq. \eqref{e5.7}:
\begin{equation}\label{e5.8f}
\frac{1}{2}\NB{y}^\nu\,\NB{g(-iy)}\sum_s{\GB{\frac{\ga(z_s)}{z_s^{\nu}}}\,\frac{1}{\NB{z_s+iy}}}\leq Me^{\,\al y}.
\end{equation}
Since $t\geq0$, the integral \eqref{e3.17} must be convergent for $t\geq0$, and 
this means that $\al=0$. Hence, the above inequality becomes:
\begin{equation}\label{e5.8g}
\NB{y}^\nu\,\NB{g(-iy)}\sum_s{\GB{\frac{\ga(z_s)}{z_s^{\nu}}}\,\frac{1}{\NB{z_s+iy}}}\leq M.
\end{equation}
There, the factor $\tfrac{1}{2}$ was absorbed into the constant $M$. A curious consequence of the condition \eqref{e5.8g} is that the form factor 
cannot be a Gaussian function, because if $g(E)=e^{-aE^2}$, where $a>0$, 
the left hand of Eq. \eqref{e5.8g} would be not bounded over the negative imaginary axis. 
Finally, notice that the form factor could be obtained by solving the 
integral equation \eqref{e5.8} subject to the 
conditions \eqref{e5.8d} and \eqref{e5.8g}. 
   
In many of the physical examples of unstable states such as radioactive 
nuclei, the poles are such that their imaginary parts are 
much less that their real parts, i.e., $\Im{z_s}\ll\Re{z_s}$. These poles are  
referred to as {\it narrow resonances}. In such cases, the 
coefficients $\ga(z_s)$ do not depend on the residues of the density of 
states and is a constant equal to $i$ 
\cite{ourpapercriticaltimes,ourpaperSmatrix}. Thus, for narrow resonances,
\begin{equation}\label{e5.8a}
\ga(z_s)= -2\,\frac{R(z_s)}{g(z_s)}=-i,\quad\Im{z_s}\ll\Re{z_s},
\end{equation} 
or
\begin{equation}\label{e5.8b}
\frac{R(z_s)}{g(z_s)}=\frac{i}{2},\quad\Im{z_s}\ll\Re{z_s}.
\end{equation}
Using Eq. \eqref{e5.8a}, Eq. \eqref{e5.7} transforms to
\begin{equation}\label{e5.8c}
\rho(E)=E^\nu\,g(E)\,\Im{\sum_{s>0}{\frac{1}{z_s^{\nu}}\,\frac{1}{z_s-E}}},\quad\Im{z_s}\ll\Re{z_s}.
\end{equation}
\section{Specific form factors}\label{S6}
We shall now discuss the advantages or disadvantages of 
choosing some particular form of the form factor in the density of 
states to obtain the survival amplitude at small times. 
Our two choices are the exponential form factor 
(commonly used for particle and nuclear resonances) 
and a constant which is used in a potential description of unstable states.

\subsection{Exponential form factor}\label{S6-a} 
\begin{figure}[htb!]
\centering
\includegraphics[scale=0.6]{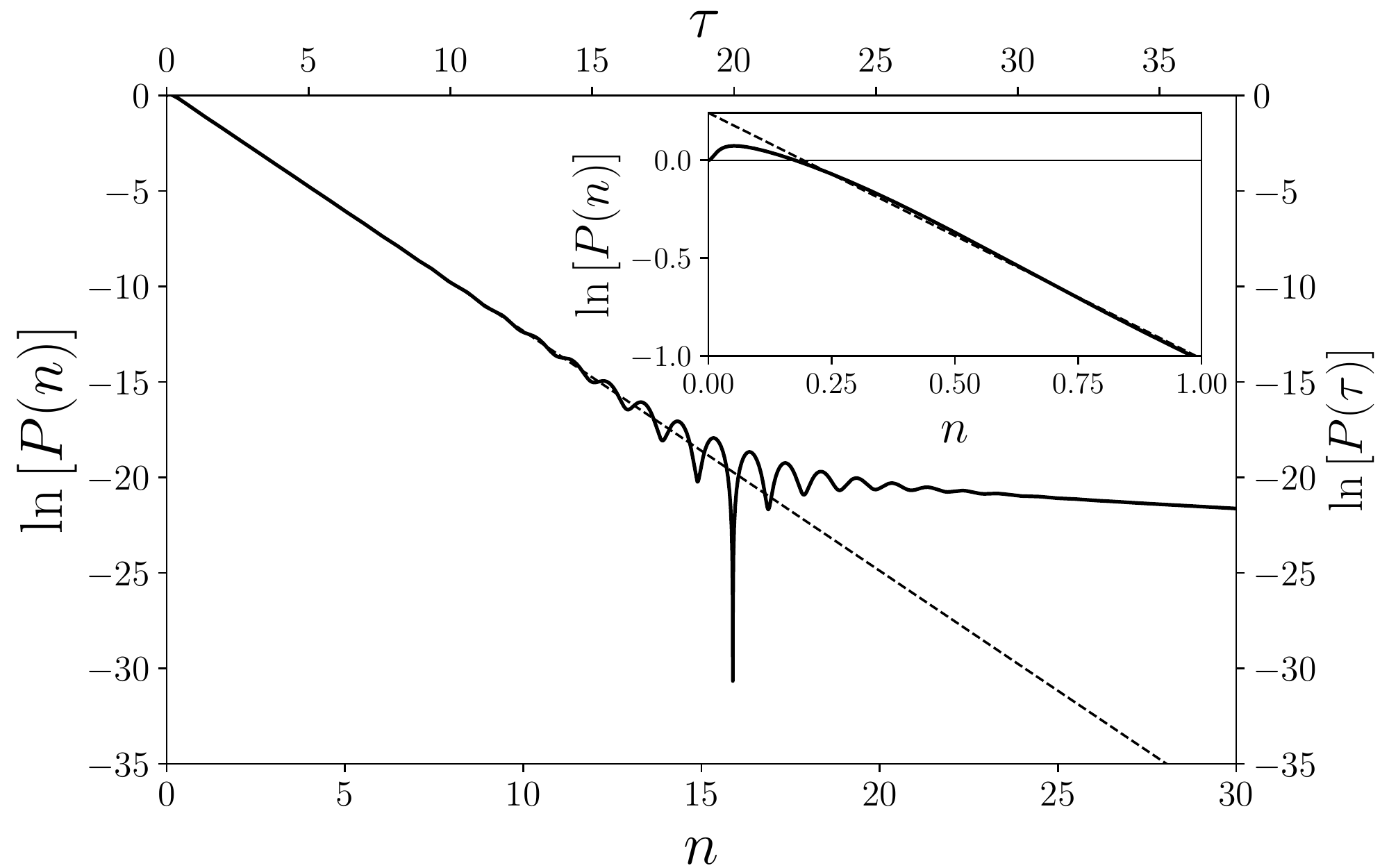}
\caption{Survival probability for a system described by an isolated resonance 
such that $x_s=0.1$, $\nu=1/2$, and $b_s=0.1$ as a function of the number 
of oscillations $n$, and the dimensionless variable for time $\tau$ given by the Eq. \eqref{e5.19}. The dashed line represents the exponential 
component of the survival probability.}\label{fig2}
\end{figure}

This is one popular choice 
in the literature because it allows closed expressions to be obtained easily. 
In this case the form factor $g(E)$ takes the form:
\begin{equation}\label{e5.9}
g(E)=e^{-bE},\quad b>0.
\end{equation}
Even if there is no physical reason for choosing this particular form, 
it satisfies the requirements imposed on the form factor, i.e., 
it ensures the convergence of the integrals \eqref{e4.7} and the condition \eqref{e5.8g} will be satisfied if $0<\nu\leq 1$. Substituting Eq. \eqref{e5.9} in 
Eq. \eqref{e5.7}, the density of states is given by 
\begin{equation}\label{e5.9a}
\rho(E)=\rho(E)=\frac{1}{2}\,E^\nu\,e^{-bE}\,{\sum_{s}{\frac{\ga(z_s)}{z_s^{\nu}}\,\frac{1}{z_s-E}}}=E^\nu\,e^{-bE}\,\Re{\sum_{s>0}{\frac{\ga(z_s)}{z_s^{\nu}}\,\frac{1}{z_s-E}}}.
\end{equation}
We shall first calculate the expectation values of the powers of the 
Hamiltonian at the initial state by substituting Eq. \eqref{e5.7} in 
Eqs \eqref{e4.7}:
\begin{equation}\label{e5.10}
\ev{H^n}_0=\intseminf{E^n\rho(E)}{E}=\Re{\sum_{s>0}{\frac{\gamma(z_s)}{z_s^{\nu}}\,\intseminf{\frac{E^{\nu+n}}{E-z_s}\,e^{-bE}}{E}}}.
\end{equation}
From the identity \cite{Erderly}
\begin{equation}\label{e5.11}
\intseminf{\frac{x^\nu}{x+\sig}\,e^{-sx}}{x}=\Ga\NP{\nu+1}\,e^{\,\sig s}\,\sig^\nu\,\Ga\NP{-\nu,\sig s},\quad \Re{s}>0,\Re{\nu}>-1,|\Arg{\sig}|<\pi,
\end{equation}
where $\Ga(\al,z)$ is the incomplete gamma function\footnote{Actually, there are two differents incomplete gamma functions: the lower and upper ones \cite{Lebedev}. Here, we are using the latter one, which is defined as
\[
\Ga(\al,z)=\int_{z}^{\infty}{t^{\al-1}e^{-t}\,dt},\quad\Re{\al}>0.
\]}, we obtain:
\begin{equation}\label{e5.12}
\ev{H^n}_0=\intseminf{E^n\rho(E)}{E}=(-1)^{n+1}\Ga\NP{1+\nu+n}\Re{\sum_{s>0}{e^{\,i\nu\pi}{z_s^n\,\ga(z_s)}\,e^{-bz_s}\,\Ga\NP{-\nu-n,-bz_s}}}.
\end{equation}
Since $\ev{H^0}_0=1$, the residues of the density of states satisfy
\begin{equation}\label{e5.13}
-\Ga\NP{1+\nu}\Re{\sum_{s>0}{e^{\,i\nu\pi}{\,\ga(z_s)}\,e^{-bz_s}\,\Ga\NP{-\nu,-bz_s}}}=1.
\end{equation}
Eq. \eqref{e5.12} must incorporate Eq. \eqref{e5.13} in order to 
ensure the normalization condition. Thus,
\begin{equation}\label{e5.14}
\ev{H^n}_0=(-1)^n\frac{\Ga\NP{1+\nu+n}}{\Ga\NP{1+\nu}}\,\frac{\Re{\sum_{s>0}{e^{\,i\nu\pi}{z_s^n\,\ga(z_s)}\,e^{-bz_s}\,\Ga\NP{-\nu-n,-bz_s}}}}{\Re{\sum_{s>0}{e^{\,i\nu\pi}{\,\ga(z_s)}\,e^{-bz_s}\,\Ga\NP{-\nu,-bz_s}}}}.
\end{equation}
In the case of narrow resonances, Eq. \eqref{e5.8a} gives us an expression 
for the residues of the density of states. Therefore, Eq. \eqref{e5.14} 
takes the following form:
\begin{equation}\label{e5.15}
\ev{H^n}_0=(-1)^n\frac{\Ga\NP{1+\nu+n}}{\Ga\NP{1+\nu}}\,\frac{\Im{\sum_{s>0}{e^{\,i\nu\pi}{z_s^n\,e^{-bz_s}}\,\Ga\NP{-\nu-n,-bz_s}}}}{\Im{\sum_{s>0}{e^{\,i\nu\pi}{\,e^{-bz_s}}\,\Ga\NP{-\nu,-bz_s}}}},\quad\Im{z_s}\ll\Re{z_s}.
\end{equation}
On the other hand, we can calculate the survival amplitude using 
the identity \eqref{e5.11}:
\begin{equation}\label{e5.16}
A(t)=\intseminf{\rho(E)e^{-iEt}}{E}=\frac{{\sum_{s}{e^{\,i\nu\pi\sgn{(s)}}{\ga(z_s)}\,e^{-pz_s}\,\Ga\NP{-\nu,-pz_s}}}}{{\sum_{s}{e^{\,i\nu\pi\sgn(s)}{\,\ga(z_s)}\,e^{-bz_s}\,\Ga\NP{-\nu,-bz_s}}}},
\end{equation}
where $p=b+it$. Even though the Eqs \eqref{e5.14} and \eqref{e5.15} allow us to compute the coefficients of the Taylor expansion, we have to choose 
a suitable value for the parameter $b$. It must be chosen such that the 
inequality \eqref{e4.6} is satisfied, 
i.e., $\ev{(\Delta H)^{2}}_0$ will be positive. Since establishing the 
range of the values of $b$ for which  $\ev{(\Delta H)^{2}}_0>0$ 
analytically is almost impossible, it is better to achieve this by 
employing numerical and graphical methods. 
In order to ilustrate this procedure, the following example is 
worked out for one isolated, narrow resonance. 
Introducing the parameters $x_s$, the normalized pole $\xi_s$, 
and the dimensionless variable for time, $\tau$, as
\begin{align}
x_s&\equiv\frac{\Im{z_s}}{\Re{z_s}},\label{e5.17}\\
\xi_s&\equiv\frac{z_s}{\Re{z_s}}=1-ix_s\sgn{(s)},\label{e5.18}\\
\tau&\equiv 2t\Im{z_s},\label{e5.19}
\end{align}
the terms $bz_s$ and $pz_s$ in Eqs \eqref{e5.14}, \eqref{e5.15} 
and \eqref{e5.16} have to be replaced by
\begin{align}
bz_s=\NP{b\Re{z_s}}\xi_s=b_s\xi_s,\label{e5.20}\\
pz_s=(b+it)z_s=\GP{b_s+i\frac{\tau}{2x_s}}\xi_s,\label{e5.21}
\end{align}
where $b_s=b\Re{z_s}$. The parameter $x_s\ll1$ indicates that a pole represents a narrow resonance, and it is related to the oscillation frequency $f_s$ of the survival probability as well, with $f_s$ given  by \cite{ourpapercriticaltimes,winter}:
$f_s=\frac{1}{4\pi x_s}$.
This frequency allows us to measure the time of the decay in terms of the 
number of oscillations $n$ that the survival probability has performed:
\begin{equation}\label{e5.23}
n=\frac{\tau}{4\pi x_s},
\end{equation} 
which is suitable for the description of the survival probability at 
small times. The choice of a suitable value of $b$ (or $b_s$) depends 
on the sign of $\ev{(\Delta H)^{2}}_0$. For $x_s=0.1$ and $\nu=1/2$, $\ev{(\Delta H)^{2}}_0$ is negative 
if $3.51\times10^{-4}<b_s<0.55$. If we choose a value of $b_s$ such that 
$\ev{(\Delta H)^{2}}_0<0$, we would expect the survival probability 
to take values greater than one. We illustrate this feature in Fig. 
\ref{fig2} by computing $P(t)$ with $b_s=0.1$. 

In this case, we can see that the survival probability is greater than one 
approximately at the first one-fourth of the oscillation. 
Since the survival probability is convex at $t=0$, 
the form factor with the value of $b_s$ chosen forces $P(t)$ 
to increase until a maximum. This behavior makes no physical sense. 

\begin{figure}[htb!]
\includegraphics[scale=0.6]{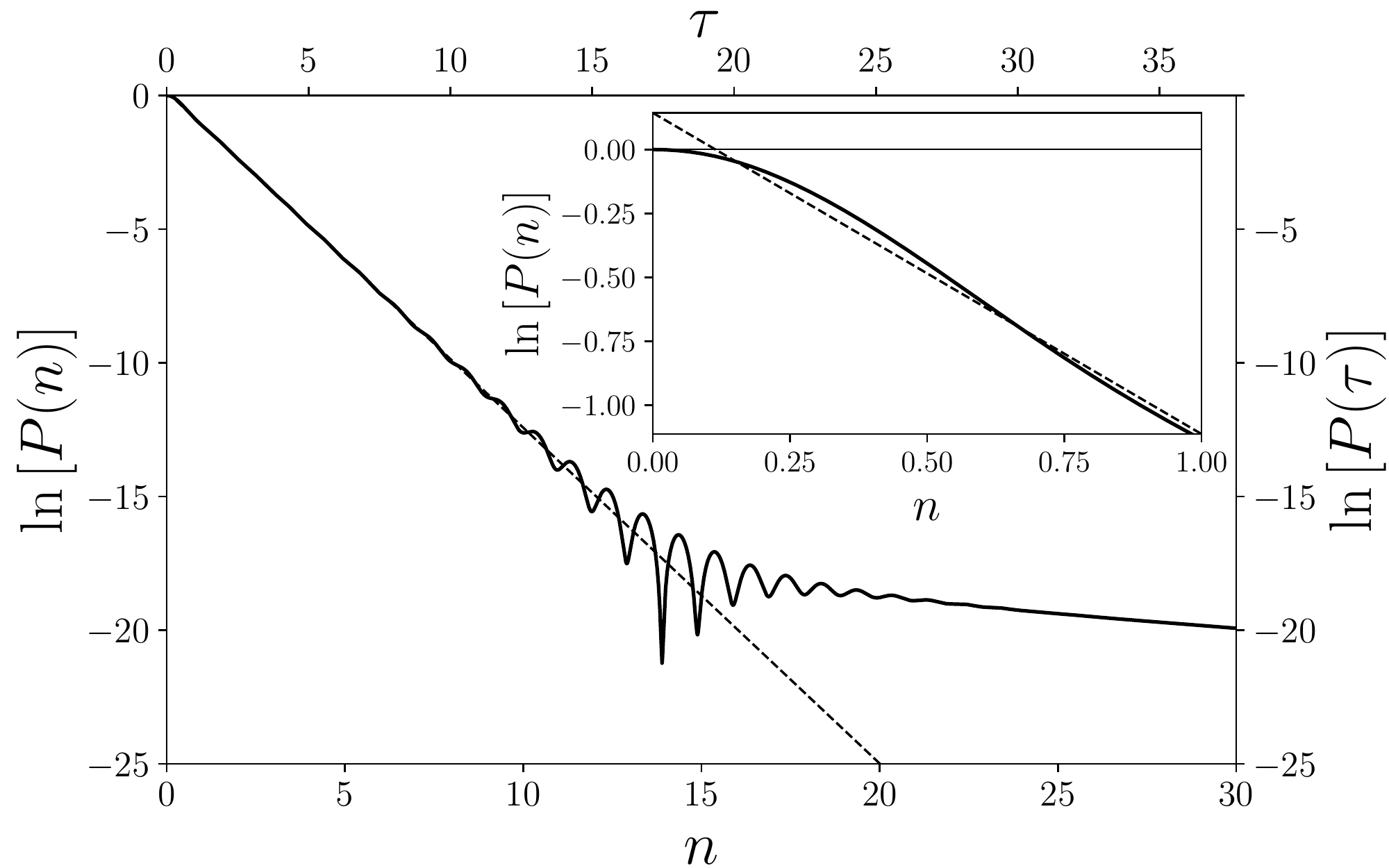}
\caption{Survival probability for a system described by an isolated resonance 
(with $x_s=0.1$, $\nu=1/2$, and $b_s=1$) as a function of the number of 
oscillations $n$, and the dimensionless variable for time $\tau$ given by the Eq. \eqref{e5.19}. The dashed line represents the exponential component of 
the survival probability.}\label{f6.3}
\end{figure}

This situation enables us to pick $b_s$ such that $\ev{(\Delta H)^{2}}_0>0$ 
for two intervals. Moreover, the larger the value of $b_s$, 
the shorter the critical time that characterizes the transition from 
the exponential to the power law behavior. 
Such a sensitivity of the critical time to the value of $b_s$ is however seen 
only for large values of $b_s$. 
If $b_s\ll1$ we do not face the problem of a variable critical time 
since the effect of the form factor over $P(t)$ is weak. 
However, the convergence of the survival probability depends 
on the condition $\Re{p}>0$, and then, a singularity is located at $p=0$. 
Hence, we should expect that the survival probability experiences strong 
variations when both $t$ and $b_s$ are close to zero. In Fig. \ref{f6.3} 
we plot the survival probability for $x_s=0.1$ and a relatively large 
value of $b_s$, i.e., $b_s=1$. Note that this value of $b_s$ ensures a 
survival probability less than one for $n\geq0$.  

\begin{figure}[htb!]
\centering
\includegraphics[scale=0.5]{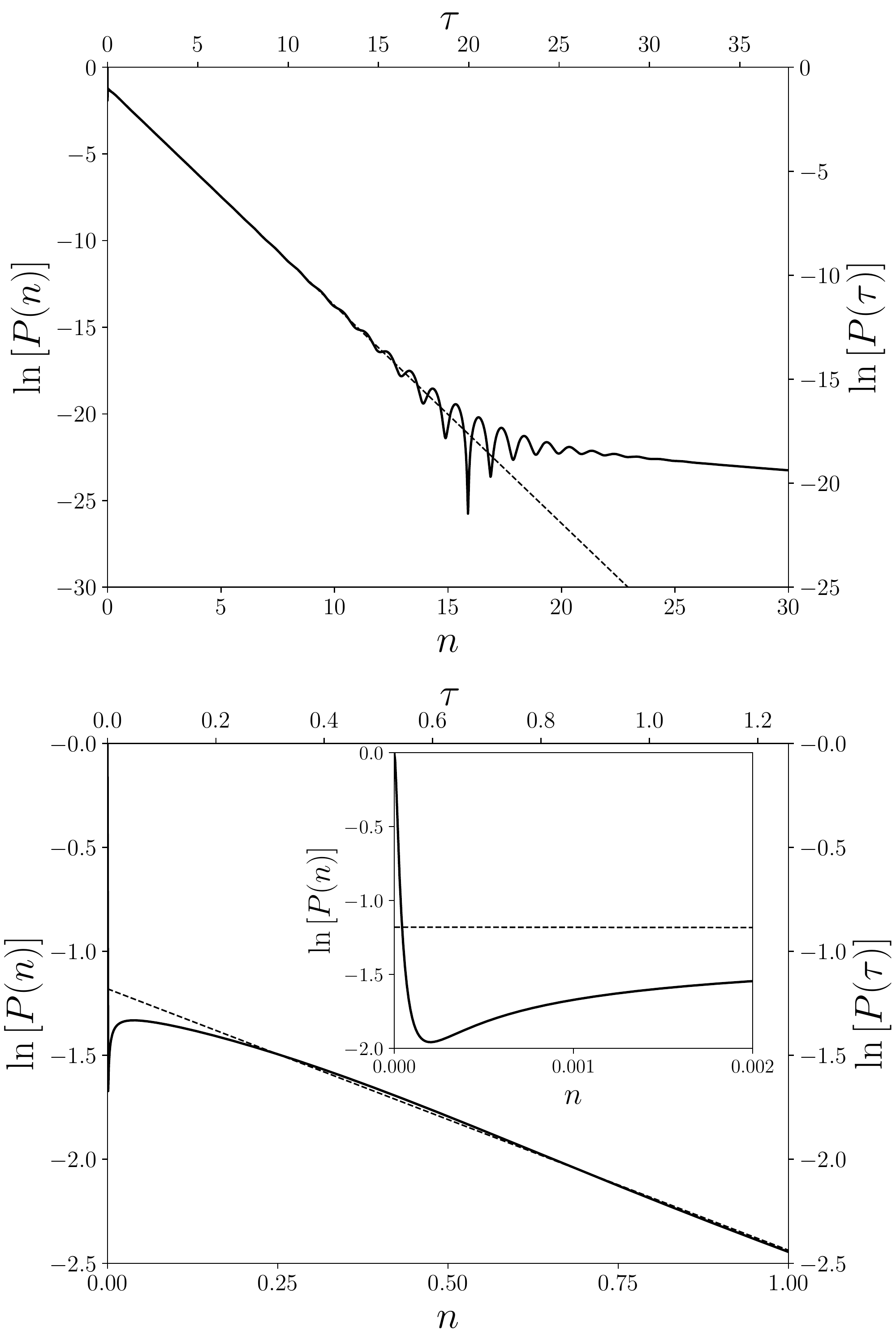}
\caption{Survival probability for a system described by an isolated resonance 
($x_s=0.1$, $\nu=1/2$, and $b_s=10^{-4}$) as a function of the number of 
oscillations $n$, and the dimensionless variable for time $\tau$ given by the Eq. \eqref{e5.19}. The upper plot shows $P(n)$ for 30 oscillations, 
and the lower one shows $P(n)$ for one oscillation. The dashed  line 
represents the exponential component of the survival probability.}\label{f6.4}
\end{figure}

Fig. \ref{f6.4} shows the survival probability for $x_s=0.1$ and 
$b_s=10^{-4}$. As expected, the survival probability is less than one 
for $n\geq0$, its critical time (for the transition from the exponential to 
the power law) has changed with respect to the previous 
example, i.e., it happens around 15 to 20 oscillations and near $t=0$, 
the survival probability decreases from 1 to approximately $e^{-2}$ in 
less than two thousandth of a period of oscillation of $P(t)$. 

\begin{figure}[htb!]
\includegraphics[scale=0.65]{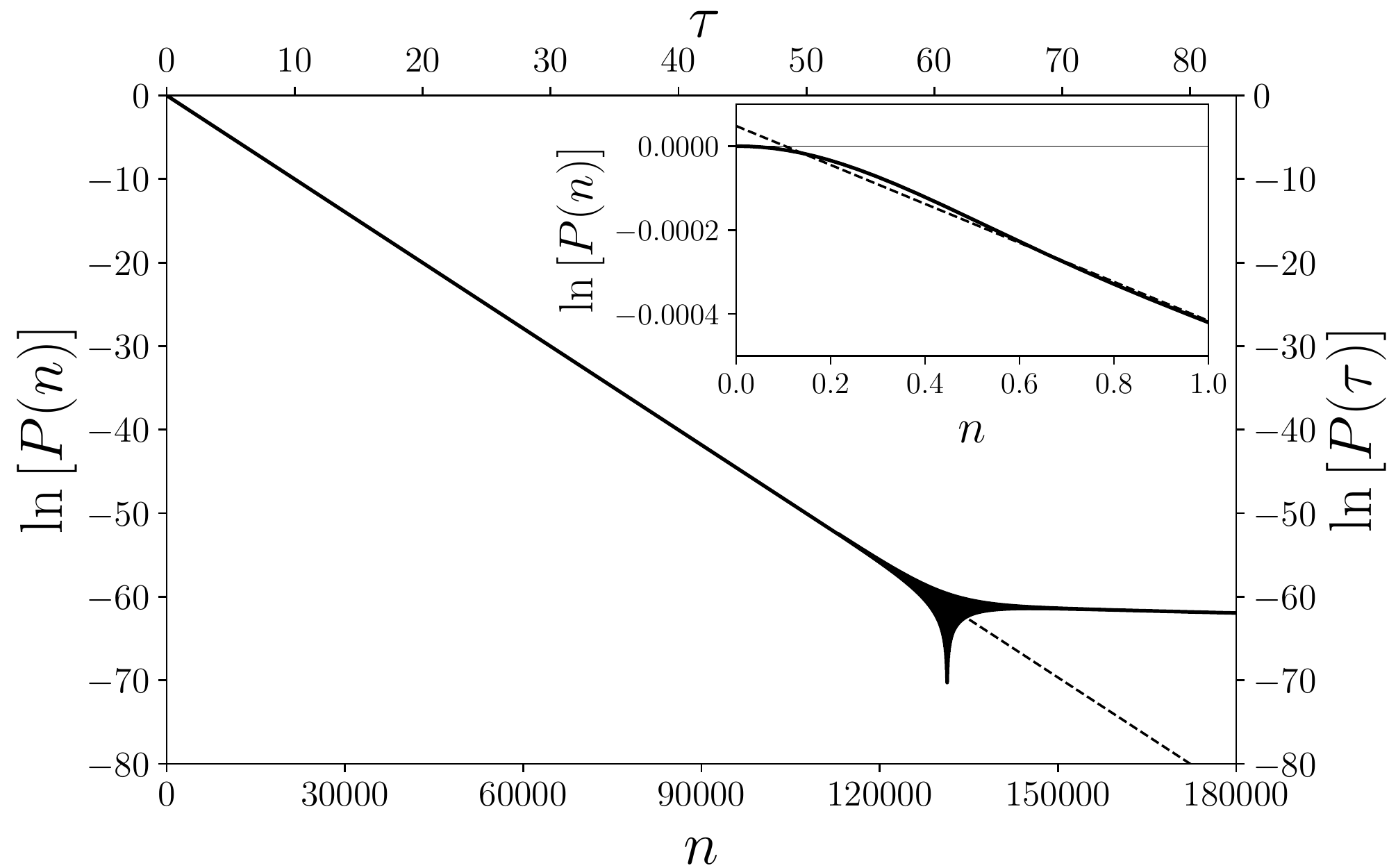}
\caption{Survival probability for the decay of $^8\text{Be}(0^+)$ into 
two alpha particles with $b_s=1$ as a function of the number of oscillations 
$n$, and the dimensionless variable for time $\tau$ given by the Eq. \eqref{e5.19}. The dashed line represents the exponential component of the 
survival probability.}\label{f6.6}
\end{figure}

Knowing how to choose a suitable value of the parameter $b$ ( or $b_s$), 
we shall now show what would happen with the survival probability for 
small times for a realistic case, namely, the decay of $^8\text{Be}(0^+)$ 
into two alpha particles for S-waves and assuming an exponential form factor. 
In this case, the real part of the pole is, $\Re{z_s}=92$ keV 
and the imaginary part is $5.6/2=2.8$ eV. 
Thus, $x_s=3\times10^{-5}$, and $\nu=1/2$. 
The survival amplitude oscillates with a period of 
$4.6\times10^{-4}$ mean lifetimes. 
%
Here, the sign of $\ev{(\Delta H)^{2}}_0$ is negative if 
$4.8\times10^{-11}<b_s<0.5$. Since a tiny value of $b_s$ does 
not describe the survival probability near $t=0$ because of the singularity 
at $p=b+it=0$, the appropiate value of $b_s$ should satisfy $b_s>0.5$. Hence, 
we choose $b_s=1$, or $b=b_s/\Re{z_s}=10.83 (\text{MeV})^{-1}$. 
For this value of $b_s$, the survival probability can be seen 
in Fig. \ref{f6.6}.

Let us look at the prominent features of this decay for the parameters chosen
above. The survival probability $P(t)$ 
is less than one for the value of $b_s$ that 
we chose, $P(t)$ reaches the large time region after 150000 oscillations 
approximately (around 70 mean lifetimes) and the intermediate region 
starts after the survival amplitude has completed one 
oscillation (in $4.6\times10^{-4}$ mean lifetimes).

Both examples point to the fact that the exponential behavior 
at small times is reached 
approximately after the survival amplitude has completed one oscillation. 
Although this could seem to be a coincidence, we shall show in section 
\ref{Sa} that the intermediate region indeed starts approximately after the 
first oscillation is completed by the survival probability for 
narrow isolated resonances.  

\subsection{Constant form factor} Even though a constant factor is the 
easiest and simplest choice, we shall show that this option implies 
some contradictions. The form factor $g(E)$ takes the form:
\begin{equation}\label{e5.24}
g(E)=1,
\end{equation}
and the density of states is given by 
\begin{equation}\label{e5.25}
\rho(E)=E^\nu\sum_s{\frac{R(z_s)}{z_s^{\nu}}\,\frac{1}{E-z_s}}.
\end{equation}
From the normalization condition, the density of states must satisfy
\begin{equation}\label{e5.26}
\intseminf{\rho(E)}{E}=\sum_{s}{\frac{R(z_s)}{z_s^{\nu}}\,\intseminf{\frac{E^\nu}{E-z_n}}{E}}=1
\end{equation}
From a mathematical point of view, the integral in Eq. \eqref{e5.26} 
converges if $-1<\nu<0$, and we have $\nu>0$ because the density of states 
must be analytic at $E=0$. Assuming that $\nu$ takes values in 
the range $0<\nu<1$, we can write Eq. \eqref{e5.26} as
\begin{equation}\label{e5.27}
\intseminf{\rho(E)}{E}=1=\sum_s{\frac{R(z_s)}{z_s^{\nu}}}\intseminf{E^{\nu-1}}{E}+\sum_{s}{\frac{R(z_s)}{z_s^{\nu-1}}\,\intseminf{\frac{E^{\nu-1}}{E-z_n}}{E}},
\end{equation}
and we can see that the integral is not finite because of the integral 
in the first term of the right side of Eq. \eqref{e5.27} unless 
\begin{equation}\label{e5.28}
\sum_s{\frac{R(z_s)}{z_s^{\nu}}}=0,\quad 0<\nu<1.
\end{equation}
From the Eqs \eqref{e5.27} and \eqref{e5.28}, we have
\begin{equation}\label{e5.29}
\intseminf{\rho(E)}{E}=1=\sum_{s}{\frac{R(z_s)}{z_s^{\nu-1}}\,\intseminf{\frac{E^{\nu-1}}{E-z_s}}{E}}=
-\frac{\pi}{\sin{\pi\nu}}\sum_s{R(z_s)\,e^{\,i\pi\nu\sgn{(s)}}}.
\end{equation}
In conclusion, if $0<\nu<1$ and because of the normalization condition, the residues and poles of the density of states have to satisfy the conditions
\begin{align}
\sum_s{R(z_s)\,e^{\,i\pi\nu\sgn{(s)}}}&=-\frac{\sin{\pi\nu}}{\pi},\label{e5.30}\\
\sum_s{\frac{R(z_s)}{z_s^{\nu}}}&=0\label{e5.31}.
\end{align}

Although the conditions \eqref{e5.30} and \eqref{e5.31} seem arbitrary, 
they appear naturally in systems under the influence of 
a central potential of finite range 
whose decay is calculated by either resonant states or Jost functions 
\cite{ourpapercriticaltimes}. 
Moreover, substituting $\rho(E)$ given by Eq. \eqref{e5.26} into 
Eq. \eqref{e4.7}, we get the expectation values of the integer powers of 
the Hamiltonian:
\begin{align}
\intseminf{E^m\rho(E)}{E}
&=\sum_s{\frac{R(z_s)}{z_s^{\nu}}\intseminf{E^{\nu}\frac{E^m}{E-z_s}}{E}}\notag\\
&=\sum_s{\frac{R(z_s)}{z_s^{\nu}}\intseminf{E^{\nu-1}\EC{\sum_{p=0}^{m}{z_s^pE^{m-p}}+\frac{z_s^{m+1}}{E-z_s}}}{E}}\notag\\
&=\sum_{p=0}^{m}{\EP{\intseminf{E^{m+\nu-p-1}}{E}}\EP{\sum_s{z_s^p\,\frac{R(z_s)}{z_s^\nu}}}}
+\sum_s{z_s^{m+1}\frac{R(z_s)}{z_s^\nu}\intseminf{\frac{E^{\nu-1}}{E-z_s}}{E}}.\label{e5.32}
\end{align} 
Eq. \eqref{e5.32} will be finite if
\begin{equation}\label{e5.33}
\sum_s{z_s^{p}\frac{R(z_s)}{z_s^\nu}}=0,\quad p=0,1,\dotsc,m.
\end{equation}
Since this condition should be valid for all positive $m$, 
it can be written as:
\begin{equation}\label{e5.34}
\Re{\sum_{s>0}{z_s^{p}\frac{R(z_s)}{z_s^\nu}}}=0,\quad p=0,1,\dotsc,m,\dotsc
\end{equation}
The expectation value of $H^n$ at the initial state is therefore given by, 
\begin{multline}\label{e5.35}
\ev{H^m}_0=\intseminf{E^m\rho(E)}{E}=\sum_s{z_s^{m+1-\nu}R(z_s)\intseminf{\frac{E^{\nu-1}}{E-z_s}}{E}}
\\=-\frac{\pi}{\sin{\pi\nu}}\sum_s{z_s^{m}R(z_s)\,e^{\,i\pi\nu\sgn{(s)}}},\quad m=0,1,2,\dotsc
\end{multline}
It is worth to point out that it is possible to ensure the convergence of 
the integrals \eqref{e4.7} by imposing conditions over the poles and residues 
of the density of states given by Eqs \eqref{e5.33} and \eqref{e5.35}. 
For $p=0$, we get the condition \eqref{e5.31} which was obtained 
by the normalization condition. As a result, the survival amplitude 
can be obtained from Eq. \eqref{a1.5} and the identity \eqref{e5.11}. 
After a short algebra, we get, 
\begin{equation}\label{e5.36}
A(t)=-2\pi i\sum_{s>0}{R(z_s)e^{-iz_st}}-\frac{\pi}{\sin{\pi\nu}}e^{-i\pi\nu}\sum_s{R(z_s)e^{-itz_s}\frac{\Ga\NP{-\nu,-iz_st}}{\Ga\NP{-\nu}}},\quad 0<\nu<1.
\end{equation}
There are some advantages to studying the decay of a system using a 
density of states with a constant form factor. Firstly, 
it lets us separate the exponential behavior from both the 
survival amplitude and probability. Secondly, it lets us deal with 
isolated resonances because the conditions \eqref{e5.30} and \eqref{e5.31} 
give us an expression for the residue in terms of its associated pole only 
(see Eq. \eqref{e6.3} below). Lastly, the descriptions of the survival 
probability in the intermediate and large time regions are simpler and 
allow us to study in detail the transition from the former to the latter 
region in a suitable way. However, using this formalism for the description 
of $P(t)$ for small times for one isolated resonance gives 
us a result that disagrees with Eq. \eqref{e4.4}. 
Let us assume that the isolated resonance has the pole, 
$z_r=\sig_r-i\om_r/2$, associated with it. 
Thus, its residue is obtained from the 
conditions \eqref{e5.13} and \eqref{e5.14} to be  
\begin{align}
\frac{R(z_r)}{z_r^\nu}+\frac{R^*(z_r)}{{z_r^*}^\nu}=0,\label{e6.1}\\
R(z_r)e^{\,i\pi\nu}+R^*(z_r)e^{-i\pi\nu}=-\frac{\sin{\pi\nu}}{\pi}.\label{e6.2}
\end{align}
Solving Eqs \eqref{e6.1} and \eqref{e6.2}, the residue is given by, 
\begin{equation}\label{e6.3}
R(z_r)=-\frac{1}{2\pi i}\,\frac{z_r^\nu\sin{\pi\nu}}{\Im{\NP{z_r^\nu\,e^{\,i\pi\nu}}}}.
\end{equation}
Since $R(z_r)$ has to satisfy the conditions \eqref{e5.16}, we get, 
\begin{equation}\label{e6.4}
\Re{\sum_{s>0}{z_s^{p-\nu}R(z_s)}}=\Re{\EC{z_r^{p-\nu}\,\frac{\sin{\pi\nu}}{2\pi i}\,\frac{z_r^\nu}{\Im{\NP{z_r^\nu\,e^{\,i\pi\nu}}}}}}=\frac{\sin{\pi\nu}}{2\pi\Im{\NP{z_r^\nu\,e^{\,i\pi\nu}}}}\Im{\NP{z_r^p}}=0,\text{ for all } p=0,1,2,\dotsc
\end{equation}
This equation can be written as, 
$\sin{\NC{p\Arg{z_r}}}=0,\text{ for all }p=0,1,2,\dotsc$. 
The only choice that satisfies 
this equation for all values of $p$ and 
for the range of values that $\Re{z_r}$ and $\Im{z_r}$ can take is:
$\Arg{z_r}=0$, 
a result that implies $\Im{z_r}=0$ and hence a resonance without a width. 
If one would still insist to continue with the description of such a resonance 
and examine the residue, one would find, 
\begin{equation}\label{e6.7}
\lim_{\Im{z_r}\to0}{R(z_r)}=-\frac{1}{2\pi i},
\end{equation}
and the density of states would be given by 
\begin{equation}\label{e6.8}
\rho(E)=\lim_{\Im{z_r}\to0}{\GP{-\frac{1}{\pi}}\Im{\frac{\NP{E/z_r}^{\nu}}{E-\Re{z_r}+i\Im{z_r}}}}.
\end{equation}
From the Plemelj-Dirac formula\footnote{The Plemelj-Dirac formula is:
\[
\lim_{\ep\to0}{\frac{1}{x'-x\mp i\ep}}=P\,\frac{1}{x'-x}\pm i\pi\delta(x'-x),
\]
where $P$ is the principal value \cite{Wyld}.}
\begin{align}
\rho(E)&=-\frac{1}{\pi}\GP{\frac{E}{\Re{z_r}}}^\nu\lim_{\Im{z_r}\to0}{\Im{\GC{P\,\frac{1}{E-\Re{z_r}}-i\pi\delta(E-\Re{z_r})}}}\notag\\
&=\GP{\frac{E}{\Re{z_r}}}^\nu\delta(E-\Re{z_r})=\delta(E-\Re{z_r}),\label{e6.9}
\end{align}
the survival amplitude reduces to 
\begin{equation}\label{e6.10}
A(t)=\intseminf{\delta(E-\Re{z_r})\,e^{-iEt}}{t}=e^{-it\Re{z_r}},
\end{equation}
and the decay probability will be equal to one, implying no decay.  
This means that it is not possible to obtain the survival probability 
for small times starting with the energy density of a system which is an
isolated resonance. 
\section{Transition Regions and Critical Times}\label{Sa}
We mentioned before that the survival probability has three well-defined 
regions: the small times region where $P(t)$ is dominantly quadratic, 
the intermediate region where $P(t)$ is approximately exponential 
and the large time region where $P(t)$ displays a power law behaviour. 
However, there is not really a sharp separation between the three regions but 
rather an oscillatory transition region from the quadratic to the exponential
and the exponential to the power law. 
Associated with these regions are the {\it critical times}, 
which indicate when the transition starts and ends or when the transition is 
happening. In this section, we shall explore these aspects.
\subsection{Critical Time and the Transition Region from Small to Intermediate Times}
\begin{figure}[hbt!]
\centering
\includegraphics[scale=1.5]{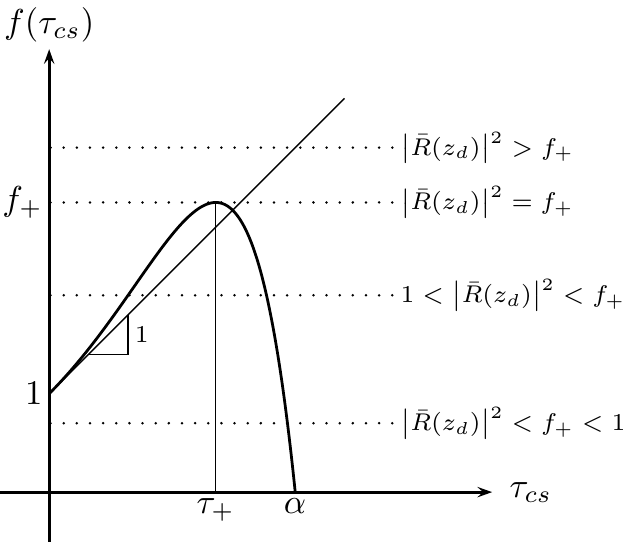}
\caption{Sketch of the function $f(\tau_{cs})$.}\label{f2}
\end{figure}

In order to find the critical time of transition, $t_{cs}$, 
from the quadratic to the exponential region we begin by 
approximating $P(t)$ given by \eqref{e4.6a}, at small times, as 
\begin{equation}\label{e7.1}
P(t)\approx1-\ev{(\Delta H)^2}_0t^2.
\end{equation} 
Focussing on the intermediate region, we note that from Eqs \eqref{e3.16} and 
\eqref{e3.21}, we know the exponential component of both the survival 
amplitude and probability. In order to simplify the notation in what follows, 
we recall that $\bar{R}(z_s)=-2\pi i{R}(z_s)$; and rewriting 
Eqs \eqref{e3.16} and \eqref{e3.21} and expressing the poles in terms of their 
real and imaginary parts, we get, 
\begin{align}
A_e(t)&=\sum_{s}{\bar{R}(z_s)e^{-i\sig_st}\,e^{\,\om_st/2}}\label{e7.2}\\
P_e(t)&=|A_e(t)|^2=\sum_{s,s'}{\bar{R}(z_s)\bar{R}^*(z_{s'})e^{-i(\sig_s-\sig_{s'})t}e^{-(\om_s+\om_{s'})t/2}}\notag\\&=\sum_{s}{|\bar{R}(z_s)|^2\,e^{-\om_st}}+\sum_{s\neq s'}{\bar{R}(z_s)\bar{R}^*(z_{s'})e^{-i(\sig_s-\sig_{s'})t}e^{-(\om_s+\om_{s'})t/2}}.\label{e7.3}
\end{align}
In order to study the transition, the simplest choice is to approximate 
$P_e(t)$ by taking the slowest decreasing term which is associated with 
the fourth-quadrant pole of the density of states having the smallest 
absolute value of the imaginary part. We call this the 
{\it decay's dominant pole}, $z_d=\sig_d-i\om_d/2$, so that 
$\om_d$ must satisfy the following: 
\begin{equation}\label{e7.4}
  \om_d=-2\min_{s}{\NL{\NB{\Im{z_s}}}}.
\end{equation}
The intermediate time survival probability can now be written approximately 
as, 
\begin{equation}\label{e54}
P(t)\approx|\bar{R}(z_d)|^2\,e^{-\om_dt}. 
\end{equation}  
The transition time $t=t_{cs}$ could be defined as the time for which 
the probabilities given by Eqs \eqref{e7.1} and \eqref{e54} are equal. 
Thus, $t_{cs}$ satisfies the equation
\begin{equation}\label{e7.6}
1-\ev{(\Delta H)^2}_0t_{cs}^2=|\bar{R}(z_d)|^2\,e^{-\om_dt_{cs}}.
\end{equation}
It is convenient to define the dimensionless quantity, 
$\tau_{cs}=\om_d t_{cs}$, which, after some algebraic manipulations 
allows us to write Eq. \eqref{e7.6} as, 
\begin{equation}\label{e7.7}
f(\tau_{cs})=\GC{1-\GP{\frac{\tau_{cs}}{\al}}^2}e^{\,\tau_{cs}}=|\bar{R}(z_d)|^2,
\end{equation}
where we have introduced the parameter $\al$ defined as,
\begin{equation}\label{e7.8}
\al=\frac{\om_d}{\sqrt{\ev{(\Delta H)^2}_0}}.
\end{equation}
The problem is thus reduced to finding the values of $\tau_{cs}$ for 
which $f(\tau_{cs})$ is equal to $|\bar{R}(z_d)|^2$. 
The next step is to study under what conditions will Eq. \eqref{e7.7} 
have a solution. Therefore, we need to study some properties of this function 
in the interval $0\leq\tau_{cs}\leq\al$ where the survival probability 
for small times is positive. 
One can easily check these properties to be: 
(i) For $\tau_{cs}=0$, $f(0)=1$ for all values of $\al$ 
and $\tau_{cs}=\al$, $f(\al)=0$.\\
(ii) Since $f'(0)=1$, the straight line which is tangent to 
$f(\tau_{cs})$ at $\tau_0$ has a slope of 1 and is independent of the value 
taken by $\al$.\\
(iii) $f(\tau_{cs})$ has a maximum in the interval $(0,\al)$ given by 
$\tau_+=\sqrt{1+\al^2}-1$, and the value of the function at that point 
is $f(\tau_+)=f_+=2\tau_+e^{\,\tau_+}/\al^2$.
it is concave. But, for $\al\geq\sqrt{2}$, it has an inflection point at 
$\tau_{cs}\equiv\tau_i=\sqrt{\al^2+2}-2$ 
and it is convex for $\tau_{cs}<\tau_i$ and concave for $\tau_{cs}>\tau_i$. 
The above features of $f(\tau_{cs})$ in the interval 
$0\leq\tau_{cs}\leq\al$ are represented graphically in the figure \ref{f2}.

The maximum of $f(\tau_{cs})$ allows us to deduce a criterion for knowing 
how many solutions Eq. \eqref{e7.7} can have:
(i) If $\NB{\bar{R}(z_d)}^2 > f_+$, there is no solution.
(ii) If $f_+=\NB{\bar{R}(z_d)}^2$, there is one solution.
(iii) If $1<\NB{\bar{R}(z_d)}^2 < f_+$, there are two solutions.
(iv) If $\NB{\bar{R}(z_d)}^2 < f_+ < 1$, there is one solution.
These cases are represented graphically in Fig. \ref{f2}. 
The primary conclusion about this definition of the transition time is that 
there exist different solutions that depend on the modulus squared of the 
residue of the density of states evaluated at the dominant pole. 
Unlike the definitions adopted for the transition time from the exponential 
to the power law behavior, there is no physical criterion for choosing one 
particular solution in this case. 
In other words, this definition of the critical time for the transition 
from the small to the intermediate time region does not provide a unique 
solution and it is not possible to choose one of them from physical arguments.

\begin{figure}[htb!]
\centering
\includegraphics[scale=0.5]{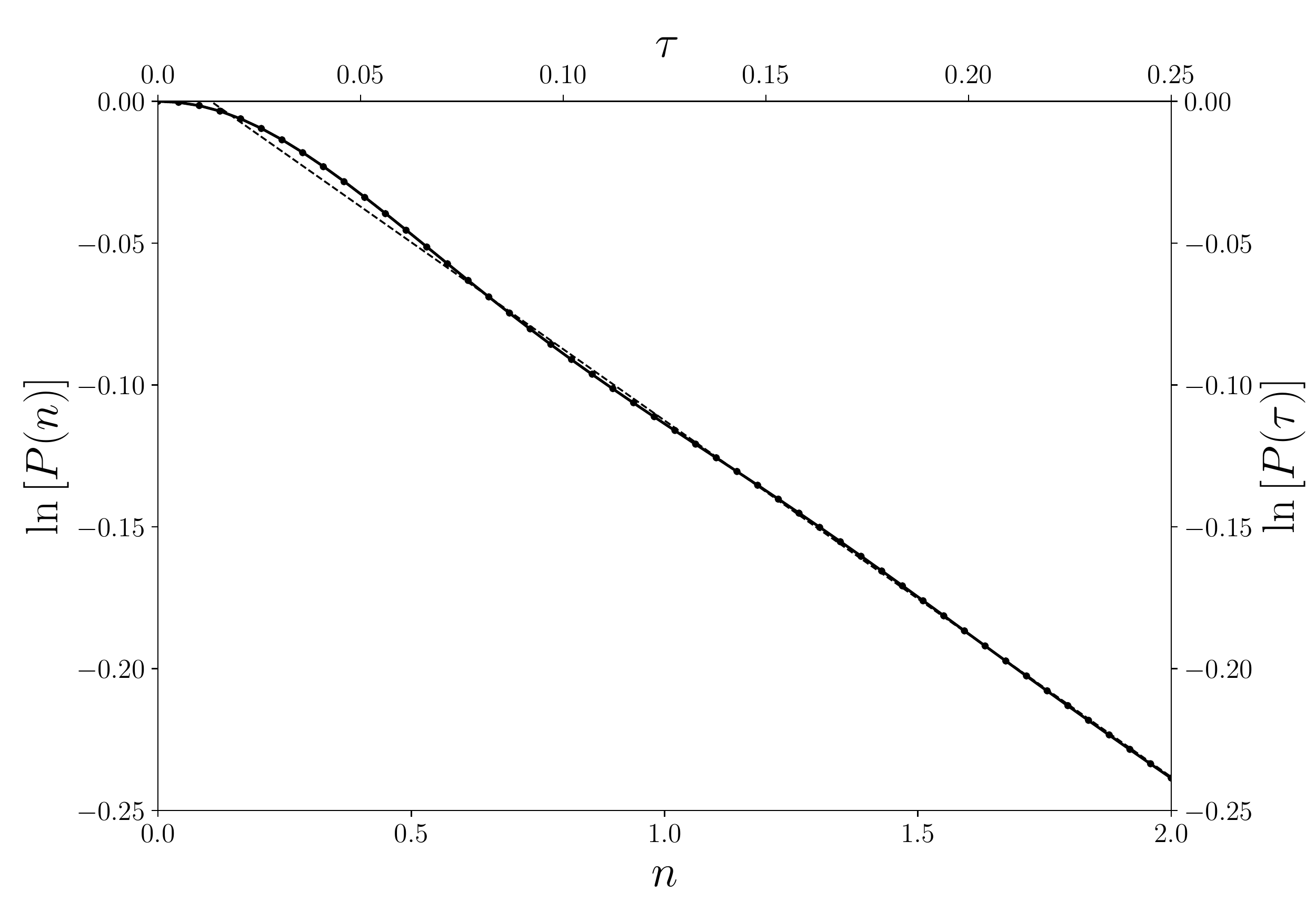}
\caption{Survival probability for small times for a system with $x_d=0.01$, and an exponential form factor with $b_s=1$ as a function of the number of oscillations 
$n$, and the dimensionless variable for time $\tau$ given by the Eq. \eqref{e5.19}. The solid line represents the complete survival probability, the dashed line represents the survival probability given by the Eq. \eqref{E7}, and the dotted line represent the exponential component of the survival probability.}\label{small}
\end{figure}

At the end of section \ref{S6-a}, we conjectured based on the examples 
developed there that the intermediate time region had been reached once the 
survival probability completes its first oscillation. 
This would mean that the modulus of the non-exponential survival amplitude at small times 
would be smaller than the modulus of the exponential survival amplitude once the survival probability completes its first 
oscillation, and this fact may give us a pointer to define 
the critical time. Furthermore, this pointer can be easily studied owing to 
the fact that the frequency of oscillations of both the survival probability and the survival amplitude are the same (which shall be shown in the section \ref{inttolarge}), which permits us to introduce the number of oscillations of the survival probability in order to analyze the temporal behavior of the exponential and non-exponential survival amplitudes. Let us then write $A_{ne}(t)$ given by 
Eq. \eqref{e3.17} and use the expression \eqref{e5.7} for the density of 
states for one isolated resonance as, 
\begin{equation}\label{e7.8-1}
A_{ne}(t)=\frac{1}{2}\sum_{s=\pm d}{\frac{\ga(z_s)}{(iz_s)^\nu}\intseminf{\frac{y^\nu}{iz_s-y}\,g(-iy)\,e^{-ty}}{y}}.
\end{equation}
Using the notation introduced by Eqs \eqref{e5.17}, \eqref{e5.18} and \eqref{e5.19} in section \ref{S6-a} for writing the time in terms of the number of 
oscillations and making the change of variable $y\to y\Re{z_s}$, we get,  
\begin{equation}\label{e7.8-2}
A_{ne}(n)=\frac{1}{2}\sum_{s=\pm d}{\frac{\ga(z_s)}{(i\xi_s)^\nu}\intseminf{\frac{y^\nu}{i\xi_s-y}\,g(-iy\Re{z_s})\,e^{-(2\pi n)y}}{y}}.
\end{equation}
Considering the case of large $n$ and writing the asymptotic expansion of 
$A_{ne}(n)$ using Watson's lemma, we get, 
\begin{equation}\label{e7.8-3}
A_{ne}(n)\sim (-i)^{\nu+1}\,g(0)\,\Re{\GP{\frac{\ga(z_d)}{\xi_d^{\nu+1}}}}\,\frac{\Ga\NP{\nu+1}}{\NP{2\pi  n}^{\nu+1}},\quad n \to \infty
\end{equation}
Note that for $n=1$, $2\pi n$ which is around 6.28 
can be considered large and thus we could say that the above equation 
is valid for $n\geq 1$. Although obtaining an analytic estimation of 
the error in using the above expression for 
all $n \geq 1$ is dificult, numerical tests validate this approximation.
From Eqs. \eqref{e3.16}, \eqref{e5.17}, \eqref{e5.18} and \eqref{e5.19}, the exponential survival amplitude can be written as, 
\begin{equation}\label{e7.8-7}
A_e(n)=-2\pi iR(z_d)\,\exp{\NP{-2\pi i\xi_d\,n}}. 
\end{equation}
Hence, 
\begin{equation}\label{e7.8-7a}
{\cfrac{A_{ne}(n)}{A_e(n)}}=\frac{1}{C(x_d,\nu)}\,\frac{e^{2\pi x_d n}}{n^{\nu+1}}\,e^{\,i2\pi n},\quad n\geq1, 
\end{equation}
where $C(x_d,\nu)$ is defined as:
\begin{equation}\label{e7.8-9}
C(x_d,\nu)=\frac{(2\pi)^{\nu+2}e^{\,i\pi\nu/2}R(z_d)}{g(0){\Ga\NP{\nu+1}}\,\Re{\NP{{\ga(z_d)}/{\xi_d^{\nu+1}}}}}.
\end{equation}
If $n\sim1$, $e^{2\pi x_dn}\ll n^{\nu+1}$ and $|A_{ne}(n)|\ll|A_e(n)|$.
This observation gets better for 
narrow resonances where $x_s\ll1$. It allows us to establish 
approximately when the small time region terminates. {\it Therefore, 
we define the critical time for the transition from the quadratic small 
time behaviour of the decay law to the intermediate exponential one as the 
time for which the survival amplitude has completed its first oscillation}. 
From Eq. \eqref{e5.23}, the critical time in the dimensionless units defined 
before is given by, 
\begin{equation}\label{e7.8-4}
\tau_{cs}=4\pi x_d.
\end{equation} 
On the other hand, we shall now provide expressions for describing not only 
the small time survival probability, but also the transition from this 
region to the exponential regime. We start by computing the ratio 
${{A_{ne}(n)}/{A_e(n)}}$ once again from the Eqs. \eqref{e7.8-2} 
and \eqref{e7.8-7}, i.e.,
\begin{align}
\frac{A_{ne}(n)}{A_{e}(n)}
&=\frac{e^{2\pi in}}{2\pi i}\sum_{s=\pm d}{\frac{\ga(z_s)}{\ga(z_d)}{\frac{1}{(i\xi_s)^\nu}}\intseminf{e^{\,2\pi x_d n}\,\frac{g\NP{-iy\Re{z_s}}}{g(z_d)}\,{\frac{y^\nu\,e^{-(2\pi n)y}}{i\xi_s-y}}\,}{y}}\notag\\
&=\frac{e^{2\pi in}}{2\pi i}\intseminf{e^{\,2\pi x_d n}\,\frac{g\NP{-iy\Re{z_d}}}{g(z_d)}\,{y^\nu\,e^{-(2\pi n)y}}\EC{{\frac{1}{(i\xi_d)^\nu}}\,\frac{1}{i\xi_d-y}+\frac{\ga(z_d^*)}{\ga(z_d)}\,{\frac{1}{(i\xi_d^*)^\nu}}\,\frac{1}{i\xi_d^*-y}}}{y}.\label{E3}
\end{align}  
Since we are dealing with narrow resonances ($x_d\ll1$) and since $n<1$, 
$\ga(z_d)\approx-i$, $\exp{\NP{\,2\pi x_d n}}\approx1$, 
and $\xi_d\approx e^{-ix_d}$. Hence, Eq. \eqref{E3} becomes:
\begin{equation}\label{E4}
\frac{A_{ne}(n)}{A_{e}(n)}\approx\frac{e^{2\pi in}}{2\pi i}\,\intseminf{\frac{g\NP{-iy\Re{z_d}}}{g(z_d)}\,(-iy)^\nu\,e^{-(2\pi n)y}\,\EC{{\frac{e^{-i\nu x_d}}{y-ie^{\,ix_d}}-\frac{e^{\,i\nu x_d}}{y-ie^{-ix_d}}}}}{y}.
\end{equation}
Without the knowledge of the specific form of the form factor, we cannot make 
more approximations. However, as $n$ goes from 0 to 1, we could expect that the integral decreses because of the exponential factor $e^{-2\pi ny}$. 
Calling this integral $N(n)$, i.e.,
\begin{equation}\label{E5}
N(n)=\frac{1}{2\pi i}\intseminf{\frac{g\NP{-iy\Re{z_d}}}{g(z_d)}\,(-iy)^\nu\,e^{-(2\pi n)y}\,\EC{{\frac{e^{-i\nu x_d}}{y-ie^{\,ix_d}}-\frac{e^{\,i\nu x_d}}{y-ie^{-ix_d}}}}}{y},
\end{equation}
we have:
\begin{equation}\label{E6}
\frac{A_{ne}(n)}{A_{e}(n)}\approx e^{2\pi in}N(n),\quad x_s\ll1,0<n<1.
\end{equation}
Both the survival probability and the transition from small to intermediate 
times can be described through Eq. \eqref{E6}. Furthermore, the survival 
probability at small times will be given by
\begin{equation}\label{E7}
P(t)=\NB{A_{e}(n)+A_{ne}(n)}^2=4\pi^2|R(z_d)|^2e^{-4\pi x_dn}\NB{1+e^{2\pi in}N(n)}^2,
\end{equation}
expression that shows the oscillatory nature of the transition. An example of 
how the transition from small to intermediate time happens can be seen in the 
figure \ref{small} for $x_d=0.01$ and an exponential form factor with $b_d=1$. 
We can see that the survival probability for small times given by the 
Eq. \eqref{E5} (dashed line) agrees with the survival probability calculated 
from the eqs. \eqref{e7.8-2} and \eqref{e7.8-7} (solid line). We also see 
that the transition ends near $n=1$ by comparing with the exponential 
survival probability (dotted line).

In the section \ref{inttolarge}, we shall use the asymptotic expansion of $A_{ne}(n)$ once again and  we shall deduce the conditions on the number of oscillations, $n$, for the 
intermediate exponential region and its transition to the power law at large 
times. 
Apart from the above observations, we can also see that since the 
survival probability is convex downward at $n=0$ and it becomes  
convex upward near $n=1/2$, the quadratic law will be valid 
approximately in the first half-cycle of the survival probability, 
and the transition from the small to the intermediate time region 
occurs in the second half-cycle.
\subsection{Comparison with other approaches}\label{compareapproaches}
Investigations of the small time behaviour of the survival probability have a 
long history with the interest being particularly enhanced by the possibility 
of the so-called quantum Zeno effect \cite{sudarshan}. 
Here we shall compare the results obtained in the present work with some 
of those in literature for the behaviour of the survival probability at short
times and its transition to the exponential behaviour at intermediate times.
Before performing such a comparison, we must emphasize that most of the 
comparisons in literature do not take into account the fact that the 
survival probability $P(t)$ must be an even function of $t$. For example, 
in \cite{sudarshan}, on the basis of models, the authors propose that as 
$t \to 0$, 
\begin{equation}\label{sudarshan}
P(t) \rightarrow 1 - \frac{\alpha }{\beta} t^{\beta}, \, \, \, \beta \ne 1.
\end{equation}
The critical time is typically determined by comparing the expansion of the 
exponential, $e^{-\gamma t}$ at small times, namely, 
$\exp(-\gamma t) = 1 \, -\, \gamma t \, +\, \frac{1}{2}\gamma ^2 t^2 + \dots$ 
with the quadratic behaviour, $P(t) = 1 - (\Delta H)^2 t^2$.
Performing such a comparison, Ghirardi {\it et al}., obtained an expression  
\cite{ghirardi1979} for the critical transition time from the quadratic to the 
exponential decay law as,
\begin{equation}
t_{cs} = \frac{2 \gamma}{2 (\Delta E)^2 + \gamma^2} 
\end{equation}
where $\Delta E$ is the uncertainty of the Hamiltonian evaluated at 
the initial state and $\gamma$ is the width of the resonance. 
The authors in \cite{ghirardi1979} 
initially obtained a result based on an 
inequality deduced by Fleming \cite{fleming} and showed that the above result 
coincided with the first one in the case of $\Delta E \gg \gamma$.   
We shall briefly describe the derivation of this result and rewrite it in 
order to compare it with the result of the present work. 

Fleming derived an inequality which provided a lower bound for the 
survival probability such that, 
\begin{equation}\label{e1}
|A(t)|\geq\cos{(t\Delta E)},\quad 0\leq t\leq\frac{\pi}{2\Delta E}\,.
\end{equation}
Assuming that the density of states is described as a narrow resonance and 
satisfies the Breit-Wigner form near the resonance, 
the critical time $\tau^G_{cs}$ was defined in \cite{ghirardi1979} 
as the intersection of the functions $\cos^2{D\tau}$ and $e^{-\tau}$. 
Hence, considering 
\begin{equation}\label{e2}
\cos^2{D\tau^G_{cs}}=e^{-\tau^G_{cs}},
\end{equation}
where $D=\Delta E/\gamma$,  
the approximate solution to Eq. \eqref{e2} given in the paper is:
\begin{equation}\label{e3}
\tau^G_{cs}=\frac{1}{D^2}.
\end{equation}
Translating Eq. \eqref{e3} to our notation: 
$\ga=\om_d$, and $\Delta E=\sqrt{\ev{(\Delta H)^2}_0}$, 
\begin{equation}\label{e4}
\tau^G_{cs}=\GP{\frac{\ga}{\Delta E}}^2=\frac{\omega_d^2}{\ev{(\Delta H)^2}_0}.
\end{equation}
The right side of the above equation is nothing but the square of the 
parameter $\al$ introduced earlier. Thus,
\begin{equation}\label{e5}
\tau^G_{cs}=\frac{\omega_d^2}{\ev{(\Delta H)^2}_0}=\al^2.
\end{equation}
In the Fock-Krylov language, the uncertainty of the energy depends on the 
density of states and hence depends on the form factor too. 
Therefore, the critical time derived in \cite{ghirardi1979} is in 
principle form factor dependent. Similarly, $\tau_{cs}$ defined by 
\eqref{e7.7} is also dependent on the form factor used. 
Our definition \eqref{e7.8-4} of the critical time, 
i.e., $\tau_{cs}=4\pi x_d$, however, 
depends only on the poles of the resonance and 
is hence independent of the form factor in the density of states.

If we rewrite this critical time in terms of $x_d=\om_d/2\sigma_d$, we get,
\begin{equation}\label{e6}
\tau^G_{cs}=\frac{4x_d^2}{\ev{(\Delta H)^2}_0/\sigma_d^2},
\end{equation}
and $\ev{(\Delta H)^2}_0\neq0$ because of the energy-time uncertainty. 
We can deduce that $\tau_{cs}\to0$ when $x_d\to0$. 
Finally, the critical time measured in terms of the 
number of oscillations is:
\begin{equation}\label{e7}
n^G_{cs}=\frac{\tau^G_{cs}}{4\pi x_d}=
\frac{x_d/\pi}{\ev{(\Delta H)^2}_0/\sigma_d^2}.
\end{equation}
\begin{table}[ht]
\centering
\begin{tabular}{|c|c|c|c||c|c|}
\cline{2-6}
\multicolumn{1}{c|}{} & \multicolumn{3}{c||}{Critical times} & \multicolumn{2}{c|}{Number of oscillations}\\\hline
$x_d$ & $\tau^G_{cs}$ 
& $\tau_{cs}$ \eqref{e7.7} & $\tau_{cs}$ (\ref{e7.8-4}) &
$n_{cs}^G$ & $n_{cs} = \tau_{cs}\eqref{e7.7}/(4\pi x_d$)\\\hline
$10^{-12}$ & $4.824\times10^{-12}$ & $1.988\times10^{-6}$ & $1.257\times10^{-11}$ & $0.384$ & $1.582\times10^{5}$  \\ \hline
$10^{-11}$ & $4.825\times10^{-11}$ & $6.287\times10^{-6}$ & $1.257\times10^{-10}$ & $0.384$ & $5.003\times10^{4}$  \\ \hline
$10^{-10}$ & $4.825\times10^{-10}$ & $1.988\times10^{-5}$ & $1.257\times10^{-9}$ & $0.384$ & $1.582\times10^{4}$  \\ \hline
$10^{-9}$ & $4.825\times10^{-9}$ & $6.287\times10^{-5}$ & $1.257\times10^{-8}$ & $0.384$ & $5.003\times10^{3}$  \\ \hline
$10^{-8}$ & $4.825\times10^{-8}$ & $1.988\times10^{-4}$ & $1.257\times10^{-7}$ & $0.384$ & $1.582\times10^{3}$  \\ \hline
$10^{-7}$ & $4.825\times10^{-7}$ & $6.287\times10^{-4}$ & $1.257\times10^{-6}$ & $0.384$ & $5.003\times10^{2}$  \\ \hline
$10^{-6}$ & $4.825\times10^{-6}$ & $1.989\times10^{-3}$ & $1.257\times10^{-5}$ & $0.384$ & $1.582\times10^{2}$  \\ \hline
$10^{-5}$ & $4.825\times10^{-5}$ & $6.291\times10^{-3}$ & $1.257\times10^{-4}$ & $0.384$ & $50.065$  \\ \hline
$10^{-4}$ & $4.826\times10^{-4}$ & $1.993\times10^{-2}$ & $1.257\times10^{-3}$ & $0.384$ & $15.858$  \\ \hline
$10^{-3}$ & $4.840\times10^{-3}$ & $6.339\times10^{-2}$ & $1.257\times10^{-2}$ & $0.385$ & $5.0446$  \\ \hline
$10^{-2}$ & $4.975\times10^{-2}$ & $0.206$ & $0.1257$ & $0.396$ & $1.639$  \\ \hline
$10^{-1}$ & $0.639$ & $0.765$ & $1.257$  & $0.509$ & $0.609$  \\ \hline
\end{tabular}
\caption{Comparison of the critical transition times, $\tau_{cs}$, 
from the non-exponential to the exponential decay 
law at small times within the approach of Ref. \cite{ghirardi1979} and that 
of the present work (Eqs \eqref{e7.7} and \eqref{e7.8-4}). 
$n_{cs}^G$ \eqref{e7} are the number of oscillations 
performed by the survival probability before reaching the exponential 
region. 
The last column gives the number $n_{cs}$ corresponding to the solution of 
\eqref{e7.7}. The number corresponding to $\tau_{cs}$ \eqref{e7.8-4} is 
always $n_{cs} = 1$.
\label{table:comparetau}
}
\end{table}
Critical times of Ref. \cite{ghirardi1979} and our Eq. \eqref{e7.7} 
for different values of $x_d = \omega_d/(2\sigma_d)$, using   
an exponential form factor with $b_s=2$ in the density of states 
for an isolated narrow resonance are compared with our form factor 
independent definition \eqref{e7.8-4} in Table \ref{table:comparetau}.
From Table \ref{table:comparetau}, 
we can infer that the critical times of Ref. \cite{ghirardi1979} are 
one order of magnitude less than the critical times given by \eqref{e7.8-4} 
whereas several orders of magnitude smaller than those of \eqref{e7.7}.  
The number of oscillations seem to be constant as long as $x_d$ is small 
but start increasing beyond $x_d$ = 0.1. 
Since the result of Ref. 
\cite{ghirardi1979} is expected to be valid for narrow resonances, a comparison 
with the above results beyond $x_d$ = 0.1 may not be appropriate. 
Considering only the cases below $x_d$ =0.1, one can draw a general conclusion
that the survival probability completes about half an oscillation 
(i.e., $n_{cs}^G \approx 0.4$) before going 
over to the dominant exponential decay law in contrast to $n_{cs} =1$ of
\eqref{e7.8-4}. However, the value of $n_{cs}^G$ is form factor dependent and 
increases with decreasing value of $b_s$. For the lowest allowed value of 
$b_s=0.65$ for example, $n_{cs}^G \approx 0.9$. 
The results given in the third and sixth column, arising from Eq. \eqref{e7.7} 
are grossly different from those of \eqref{e6} and \eqref{e7.8-4}. 
One should be cautious while drawing conclusions from the results 
of \eqref{e7.7} since 
(i) values of $\tau_{cs}$ obtained using Eq. \eqref{e7.7} are strongly form factor 
dependent (for example, there is no solution for an exponential form factor 
with $b_s =1$ even if it falls in the allowed range of 
values of $b_s$ and gives 
reasonable results for $\tau_{cs}^G$) (ii) the solution of \eqref{e7.7} is 
not always unique and there are cases as mentioned above when there exists 
no solution at all. As a consequence, \eqref{e7.8-4} which depends solely 
on the pole value of the resonance, should be considered as the reliable 
estimate of $\tau_{cs}$ of the present work.

A brief comment about the increase in the number of oscillations 
as well as the existence of a bigger exponential region in $P(t)$ with 
decreasing $x_d$ is in order here. 
These features depend on the coupling constant in the decay (see 
\cite{pascual} for a demonstration with a nice model). 
Explicit examples of strong and weak decays confirm the same \cite{bogda}.   
Taking the ratio $x_d$ to the extremes, a large value of $x_d$ implies no 
exponential decay at all for the resonance 
(see for example \cite{ournonexpo2} 
for the case of the broad $\sigma$ resonance) and a tiny $x_d$ implies an 
exponential decay for a large region of $t$. Though one cannot use the 
idea of a coupling constant for tunneling decays in nuclear physics such
as that of $^8$Be to two $^4$He nuclei, one can see in Fig. \ref{f6.6} 
that with $x_d$ = 3 $\times$ 10$^{-5}$, the survival probability displays an 
extremely large exponential region and a huge number of oscillations as 
expected. 
 
To complete the comparison, we finally note that Fleming \cite{fleming} 
defined the lifetime of the resonance as 
$\tau = \int_0^{\infty} P(t) dt$ which is the value of the autocorrelation 
function ${\cal R}(y)$ of the present work (see Eq. \eqref{e3.10}) at $y =0$. 
\subsection{Critical time and the transition region between intermediate and large times}\label{inttolarge}
The intermediate time region is characterized by its dominant exponential 
nature and in the same way, the large time region is ruled by a strong 
power law. 
One must however take into account the oscillatory nature of 
the survival probability too. The oscillatory nature has given rise to 
a lot of debate in literature \cite{ourpapercriticaltimes,winter,GCoscillate,FondaGhirardi1972}.  
Our aim in this section is to obtain expressions 
that incorporate the oscillatory feature into the description of these 
regions for systems described by one isolated resonance.

\begin{figure}[!htb]
\centering
\includegraphics[scale=0.5]{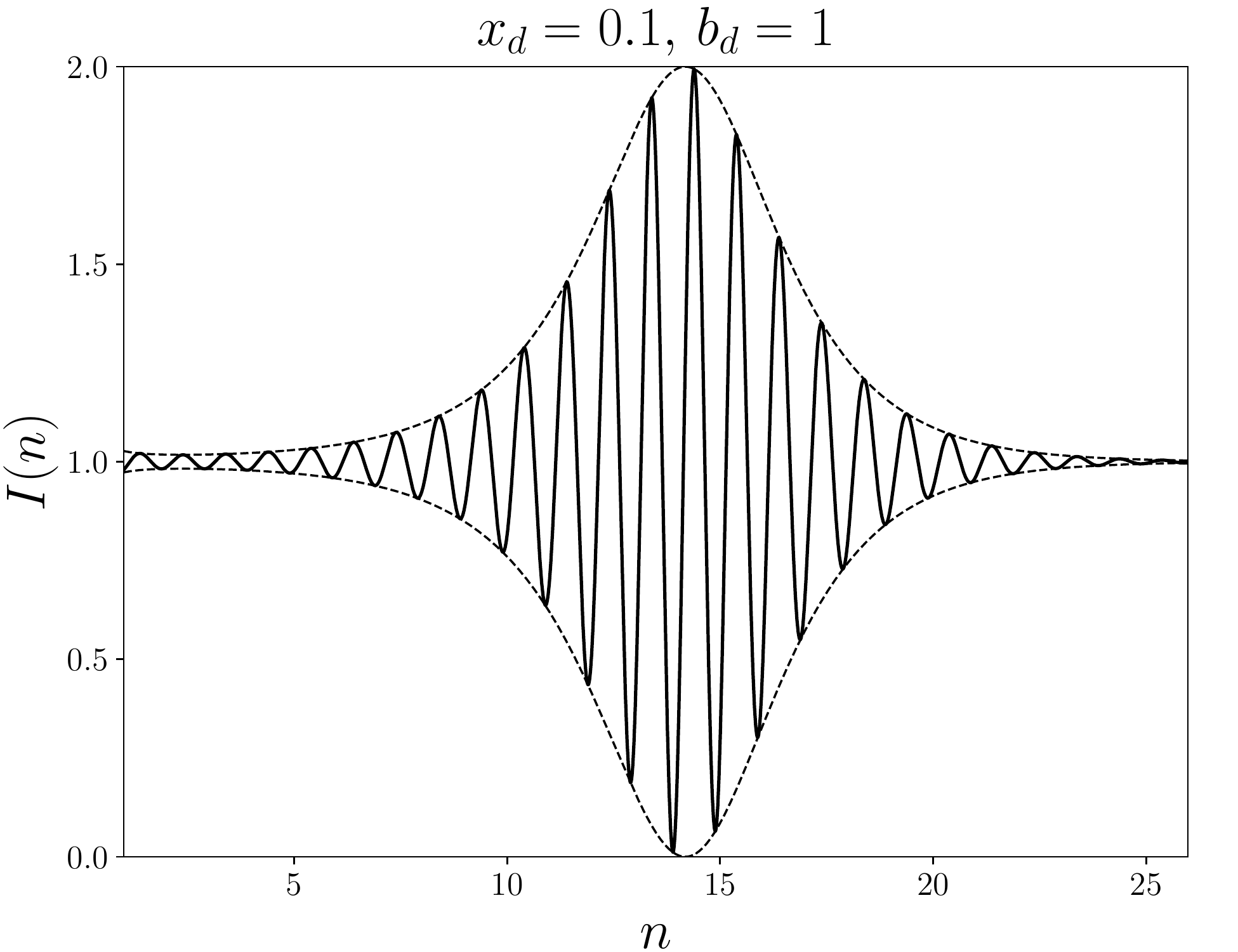}
\caption{Modulating function given by Eq. \eqref{e7.8-8} as a function of the number of oscillations $n$. Here, $x_d=0.1$, $\nu=0.5$ and an exponential form factor with $b_d=1$ 
is used.}\label{f4}
\end{figure}

In order to do this, we will use the ideas and formalism described 
in \cite{ourpapercriticaltimes} for studying the transition from  
the intermediate to the large time region. 
In ref. \cite{ourpapercriticaltimes}, the survival amplitude given by 
Eq. \eqref{e2.3} is written as the product of the sum of the 
exponential and non-exponential survival probabilities, and a modulating 
function $I(t)$:
\begin{equation}\label{e7.8-5}
P(t)=\NC{P_e(t)+P_{ne}(t)}I(t),
\end{equation}
where
\begin{equation}\label{e7.8-6}
I(t)=1+\cfrac{2\,\Re{\EC{\cfrac{A_e(t)}{A_{ne}(t)}}}}{1+\EB{\cfrac{A_e(t)}{A_{ne}(t)}}^2}=1+\cfrac{2\,\cos{\EC{\Arg{\cfrac{A_e(t)}{A_{ne}(t)}}}}}{\EB{\cfrac{A_e(t)}{A_{ne}(t)}}+\EB{\cfrac{A_{ne}(t)} {A_e(t)}}}. 
\end{equation}
Defining the function $I(t)$ seems to be the appropriate path  
for incorporating the oscillatory component in the intermediate and large time regions. To begin with, notice that the oscillation of $P(t)$ comes from the term
\[
\cos{\EC{\Arg{\cfrac{A_e(t)}{A_{ne}(t)}}}},
\]
and, from the Eqs. \eqref{e3.16} and \eqref{e3.17}, that the oscillation of the last terms comes from the exponential component of the survival amplitude, mainly, from the term $e^{-iz_dt}=e^{-i\tau/2x_d}e^{-\tau/2}$. Thus, the survival amplitude oscillates with a frequency equal to 
\[
f_d=\dfrac{1/2x_d}{2\pi}=\frac{1}{4\pi x_d},
\]
and since
\[
\cos{\EC{\Arg{\cfrac{A_e(t)}{A_{ne}(t)}}}}\propto\cos{\GP{\frac{1}{2x_d}\tau+\delta}},
\]
where $\delta$ is a constant, $I(t)$ oscillates at the same frequency as well, and so does the survival probability.  
On the other hand, substituting Eq. \eqref{e7.8-7a} in Eq. \eqref{e7.8-6}, we get, 
\begin{equation}\label{e7.8-8}
I(n)=1+\frac{2\cos{\NC{2\pi n-\Arg{C(x_d,\nu)}}}}{|C(x_d,\nu)|n^{\nu+1}\exp{\NP{-2\pi x_d\,n}}+|C(x_d,\nu)|^{-1}n^{-\nu-1}\exp{\NP{2\pi x_d\,n}}},\quad n>1,
\end{equation}
In Fig. \ref{f4} we show a plot of $I(n)$ as a function of $n$ with 
$x_d=0.1$, $\nu=0.5$, and using an exponential form factor with $b_d=1$. 

\begin{figure}[!htb]
\centering
\includegraphics[scale=0.7]{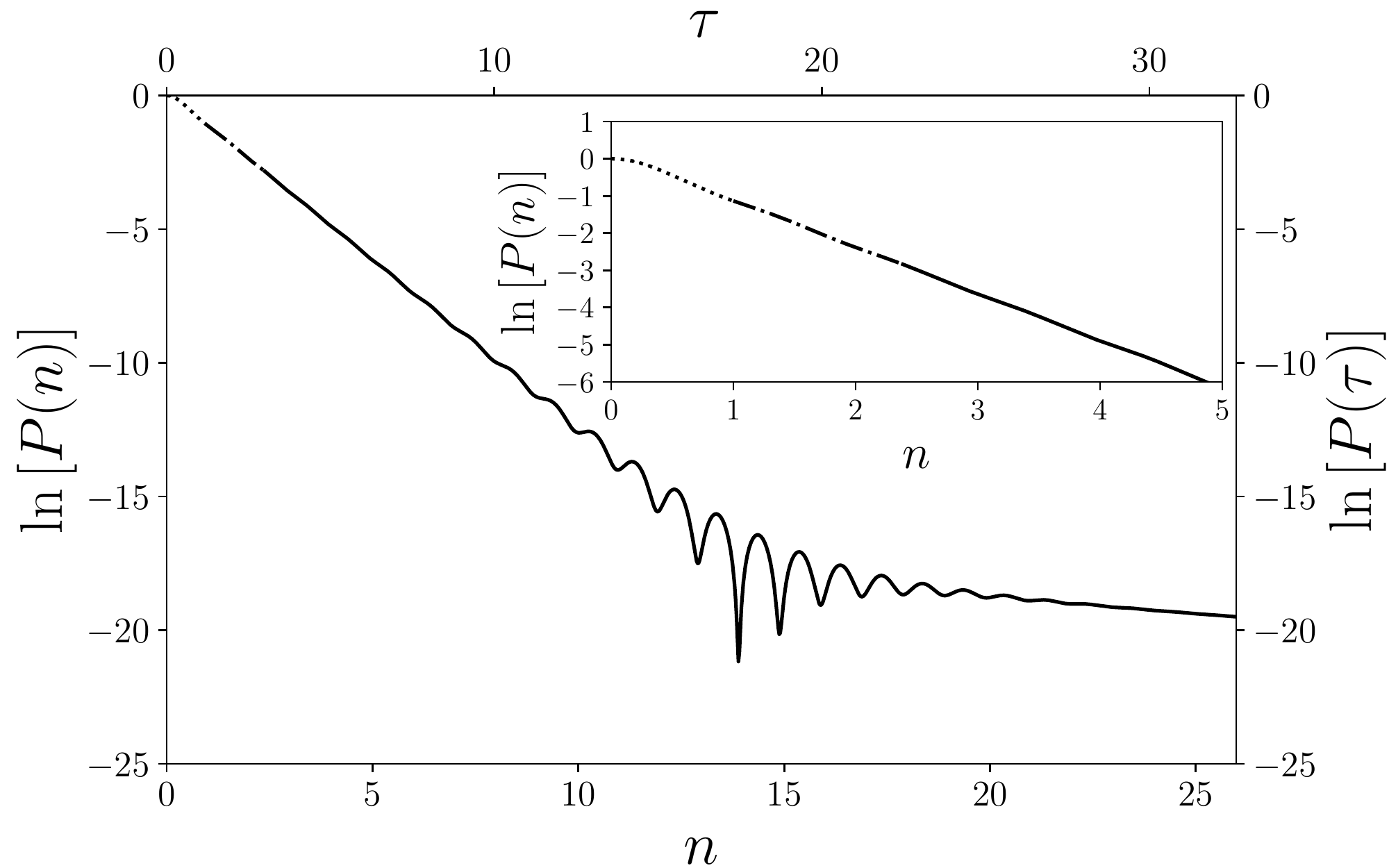}
\caption{Survival probability with parameters $x_d=0.1$, $\nu=0.5$, and an exponential form factor with $b_s=1$ as a function of the number of oscillations $n$, and the dimensionless variable for time $\tau$ given by the Eq. \eqref{e5.19}. The dotted line is the small time region, the dash-dotted line is
the intermediate time region, the solid line is the transition from the intermediate to large time region, and the dashed line is large time region.}\label{f5}
\end{figure}

For the intermediate time region, the term $|C(x_d,\nu)|^{-1}n^{-\nu-1}\exp{\NP{2\pi x_d\,n}}$ can be neglegible with respect to the term 
$|C(x_d,\nu)|n^{\nu+1}\exp{\NP{-2\pi x_d\,n}}$ and $P_e(n)\gg P_{ne}(n)$, and,
On the other hand, the large time region satisfies the opposite conditions\footnote{A demostration of this property can be seen in the appendix \ref{app2}.}. 
From Eqs \eqref{e7.8-5} and \eqref{e7.8-8}, the survival probability for 
the former region is given by
\begin{equation}\label{e7.8-10}
P(n)=4\pi^2|R(z_s)|^2\exp{\NP{-4\pi x_d\,n}}\,\EL{1+\frac{2}{|C(x_d,\nu)|}\,\frac{\exp{\NP{2\pi x_d\,n}}}{n^{\nu+1}}\cos{\NC{2\pi n-\Arg{C(x_d,\nu)}}}},
\end{equation}
and the survival probability for the latter region is given by:
\begin{equation}\label{e7.8-11}
P(n)=\frac{4\pi^2|R(z_s)C(x_d,\nu)|^2}{n^{2\nu+2}}\EL{1+{2}{|C(x_d,\nu)|}\,\frac{n^{\nu+1}}{\exp{\NP{2\pi x_d\,n}}}\cos{\NC{2\pi n-\Arg{C(x_d,\nu)}}}}.
\end{equation}
Although Eqs \eqref{e7.8-10} and \eqref{e7.8-11} incorporate 
the oscillatory nature of the survival probability, we do not have 
characteristic times that allow us to say when the intermediate or 
large time region starts or ends, or when the transition is happening. 
The first step for establishing these times is to define the critical 
time (in terms of the number of oscillations) $n_{cl}$ of the transition 
from the intermediate to large times. This is given by 
the intersection of the exponential and non-exponential survival probability \cite{ourpapercriticaltimes,bogda,garciaisolated}:
\begin{equation}\label{e7.8-12}
\NB{A_e(n_{cl})}^2=\NB{A_{ne}(n_{cl})}^2\quad\therefore\quad n_{lc}^{2\nu+2}\exp{\NP{-4\pi x_d\,n_{lc}}}=\frac{1}{|C(x_d,\nu)|^2}.
\end{equation}
This equation has two solutions and the critical time is defined as the 
largest one. Another property of this time is related with the nature of 
the resonance, i.e., the narrower the resonance is, 
the larger is the critical time\footnote{Our aim is not to discuss the 
details or demostrate the properties of the critical time for the 
transition from the intermediate to the large time region. 
In \cite{ourpapercriticaltimes}, the authors present a complete analysis 
of this for the case $\nu=1/2$, and these properties are similar for the 
range of values that $\nu$ can take here, so that the generalization 
is straightfoward.}. The next step is to analyze the critical points of the 
function
\begin{equation}\label{e7.8-13}
m(n)=\frac{2}{|C(x_d,\nu)|n^{\nu+1}\exp{\NP{-2\pi x_d\,n}}+|C(x_d,\nu)|^{-1}n^{-\nu-1}\exp{\NP{2\pi x_d\,n}}}.
\end{equation}
It is simple to verify that, for $n>1$, $m(n)$ has a maximum at $n=n_{cl}$, 
and has a minimum at $n=\frac{\nu+1}{2\pi x_d}$, or in dimensionless units, 
$\tau=2(\nu+1)$. These two times let us define where the intermediate, 
the transition and the large time regions are. 
The first interval $1<n<\frac{\nu+1}{2\pi x_d}$ corresponds to the 
intermediate time region. 
The second interval $\frac{\nu+1}{2\pi x_d}<n<2n_{cl}-\frac{\nu+1}{2\pi x_d}$ 
corresponds to the transition region: it is defined such that the critical 
time is the middle point of the interval. Finally, the interval 
$n>2n_{cl}-\frac{\nu+1}{2\pi x_d}$ corresponds to the large time region. Summarizing:
\begin{enumerate}[i)]
\item The intermediate time region is defined in 
the interval $1<n<\frac{\nu+1}{2\pi x_d}$, and the survival probability 
is given by Eq. \eqref{e7.8-10}. 
\item The transition region from the intermediate to the large time region 
is defined in the interval $\frac{\nu+1}{2\pi x_d}<n<2n_{cl}-\frac{\nu+1}{2\pi x_d}$, and the survival probability is given by
\begin{equation}\label{e7.8-14}
P(n)=I(n)\EC{4\pi^2|R(z_s)|^2\exp{\NP{-4\pi x_d\,n}}+\frac{4\pi^2|R(z_s)C(x_d,\nu)|^2}{n^{2\nu+2}}},
\end{equation}
where $I(n)$ must be given by Eq. \eqref{e7.8-8}.
\item The large time region is defined in the interval $n>2n_{cl}-\frac{\nu+1}{2\pi x_d}$, and the survival probability is given by Eq. \eqref{e7.8-11}.
\end{enumerate}
In Fig. \ref{f5}, we show a plot of $P(n)$ as a function of $n$ with 
$x_d=0.1$, $\nu=0.5$ and using an exponential form factor with $b_d=1$. 
The diferent regions have been identified by different linestyles. 

\section{Summary and Conclusions}
The time evolution of an unstable state is investigated by studying the 
properties and behaviour of the survival probability, $P(t)$, 
of a decaying state at all times. 
It is well known that the quantum mechanical description of $P(t)$ leads 
to five distinct regions, namely, the quadratic form at small times, the 
dominant exponential decay at intermediate times, the power law at large times 
and the two transition regions between the exponential and non-exponential 
decay laws at small and large times. 
Even though the topic as such has been 
studied in quite detail over decades, digging deeper into the behaviour 
of $P(t)$ in the five regions allowed us to find some subtle features as 
well as restrictions in the form of $P(t)$ and the input density of 
states (DOS) used to calculate $P(t)$. Working within the commonly used 
Fock-Krylov formalism and taking into account the conditions derived, 
each of the above mentioned five regions of the survival probability are 
studied mathematically as well as numerically with physical examples. 
Some of the main observations and conclusions are listed below: 
\begin{enumerate}[i)]
\item In the FK formalism, the survival amplitude, $A(t)$ is written as a 
Fourier transform of the DOS. Since the non-exponential decay regions are 
hard to observe and the exponential decay law dominates the observations in 
the real world, it would be useful to construct a DOS leading to a purely 
exponential decay. We show that the latter is not possible. 
\item The survival probability and the autocorrelation function of the 
density of states are shown to be a pair of cosine Fourier transforms. This 
is a particular case of the Wiener-Khinchin theorem.
\item 
A consequence of the previous result is the evenness of $P(t)$. 
\item A functional form of the DOS which depends on the pole values, a 
threshold factor and an energy dependent form factor is provided. 
Consideration of particular form factors with physical examples of resonances 
leads to the observations that (i) a constant form factor leads to unphysical 
values of the energy uncertainty, (ii) the mathematical 
condition \eqref{e5.8g} derived in this work does not allow the usage of 
a Gaussian form factor and (iii) the commonly used exponential form factor 
can be used with restrictions on the parameter in the exponential.
\item Transition regions from the small time quadratic law to the exponential 
decay law and from the exponential to the large time power law are studied 
using the constructed DOS and the mathematical conditions derived in this work.
Expressions for the critical times of transition are provided and compared 
with existing literature.
\item An interesting novel feature introduced in this work is the description 
of the survival amplitude and hence survival probability in terms of the 
number of oscillations performed. The small time quadratic behaviour is found 
to go over to the exponential decay when the survival probability has 
completed one oscillation. Analytical expressions for the 
number of oscillations performed in each of the regions of $P(t)$ are 
given in terms of the pole values and the threshold factor in the DOS. 
\end{enumerate}

An insight into the mathematical construction of the survival 
probability alongwith the different constraints and implications of the 
conditions derived for the behaviour of $P(t)$ over the entire 
region of its evolution from small to large times is thus provided in the 
present work within the Fock-Krylov framework. Since the time evolution 
of an unstable state is often studied in literature by focussing on 
a particular region of $P(t)$ or even a particular aspect of its 
behaviour, such a complete mathematical evaluation should prove useful for 
the focussed studies in future.   
\appendix
\section{Descomposition of the survival amplitude as a sum of exponential functions and an integral}\label{app1}
Consider the complex integral 
\begin{equation}\label{a1.1}
\oint_C{\rho(z)e^{-ixt}\,dz},
\end{equation}
where $C$ is the contour of integration shown in Fig. \ref{fig1}. Since $\rho(z)$ satisfies the Jordan's lemma, using the residue theorem we get, 
\begin{equation}\label{a1.2}
\int_{AO}+\int_{OB}=2\pi i\sum_{s}{e^{-iz_st}R(z_s)},
\end{equation}
However, for $AO$, $z=x$. Therefore,
\begin{equation}\label{a1.3}
\int_{AO}=-\intseminf{\rho(x)e^{-ixt}}{x}=-A(t),
\end{equation} 
and for $OB$,
\begin{equation}\label{a1.4}
\int_{OB}=-\int_{-i\infty}^{0}{\rho(z)e^{-izt}\,dz}.
\end{equation} 
The survival amplitude can be written as:
\begin{equation}\label{a1.5}
A(t)=-2\pi i\sum_{s}{e^{-iz_st}R(z_s)}-\int_{-i\infty}^{0}{\rho(z)e^{-izt}\,dz},
\end{equation}
\section{Ratio of the exponential and non-exponential survival amplitude}\label{app2}
Our aim is to show that $|A_e(n)|\gg|A_{ne}(n)|$ in the exponential region 
and $|A_e(n)|\ll|A_{ne}(n)|$ for large times. Let $\eta(n)$ be the function
\[
\eta(n)=\GB{\cfrac{A_e(n)}{A_{ne}(n)}}=|C|n^{\nu+1}e^{-2\pi x_d n}.
\]
Taking the derivative with respect to $n$ and calculating the possible critical points:
\[
\eta'(n)=\eta(n)\GP{\frac{\nu+1}{n}-2\pi x_d}=0,\quad\therefore\quad n=\frac{\nu+1}{2\pi x_d}.
\]
It is easy to show that this critical point is a maximum. Since $\eta(0)=0$ and $\eta(\infty)=0$, this function starts to rise up until a maximum, and then 
starts falling down.

These properties allow us to study the function $1/\eta(n)$, i.e., $\GB{\dfrac{A_{ne}(n)}{A_e(n)}}$. In this case, when $n$ increases from zero, 
$1/\eta(n)$ falls down until $n=(\nu+1)/(2\pi x_d)$, and then it starts to rise up indefinitely.
There should be some values of $n$ such that $\eta(n)=\dfrac{1}{\eta(n)}$. 
The values of $n$ where this condition is valid are nothing but the 
critical points for the transition from the exponential to the power law.
Since we are interested in narrow resonances, we know that the second solution 
of Eq. \eqref{e7.8-12}   
goes to zero when $x_d$ goes to zero too. This implies that even for 
$n=1$ $\eta(n)\gg\dfrac{1}{\eta(n)}$.
In conclusion, $\eta(n)>\dfrac{1}{\eta(n)}$ for $1<n<n_{cl}$, and $\eta(n)<\dfrac{1}{\eta(n)}$ for $n>n_{cl}$, where $n_{cl}$ is the critical point for 
large time transition.

\begin{figure}
\includegraphics[scale=0.5]{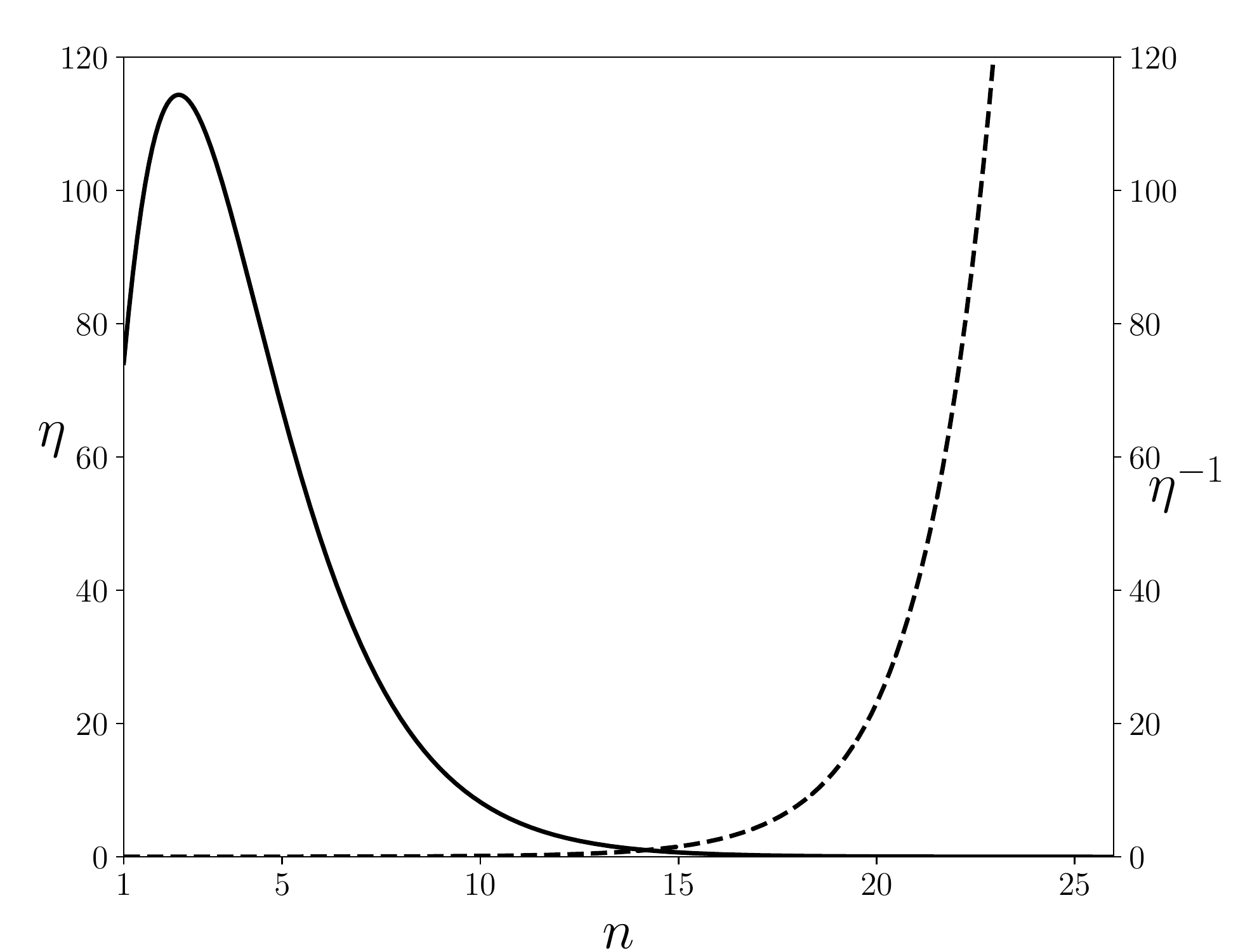}
\caption{Ratio of the exponential and non-exponential survival amplitudes, 
$\eta$ and its inverse as a function of the number of oscillations $n$ of the 
survival probability.} \label{etaplot}
\end{figure}

In Fig. \ref{etaplot}, we plot $\eta$ (solid line) and $1/\eta$ (dashed line) for $x_d=0.1$ and $b_s=1$. 

\begin{figure}
\includegraphics[scale=0.5]{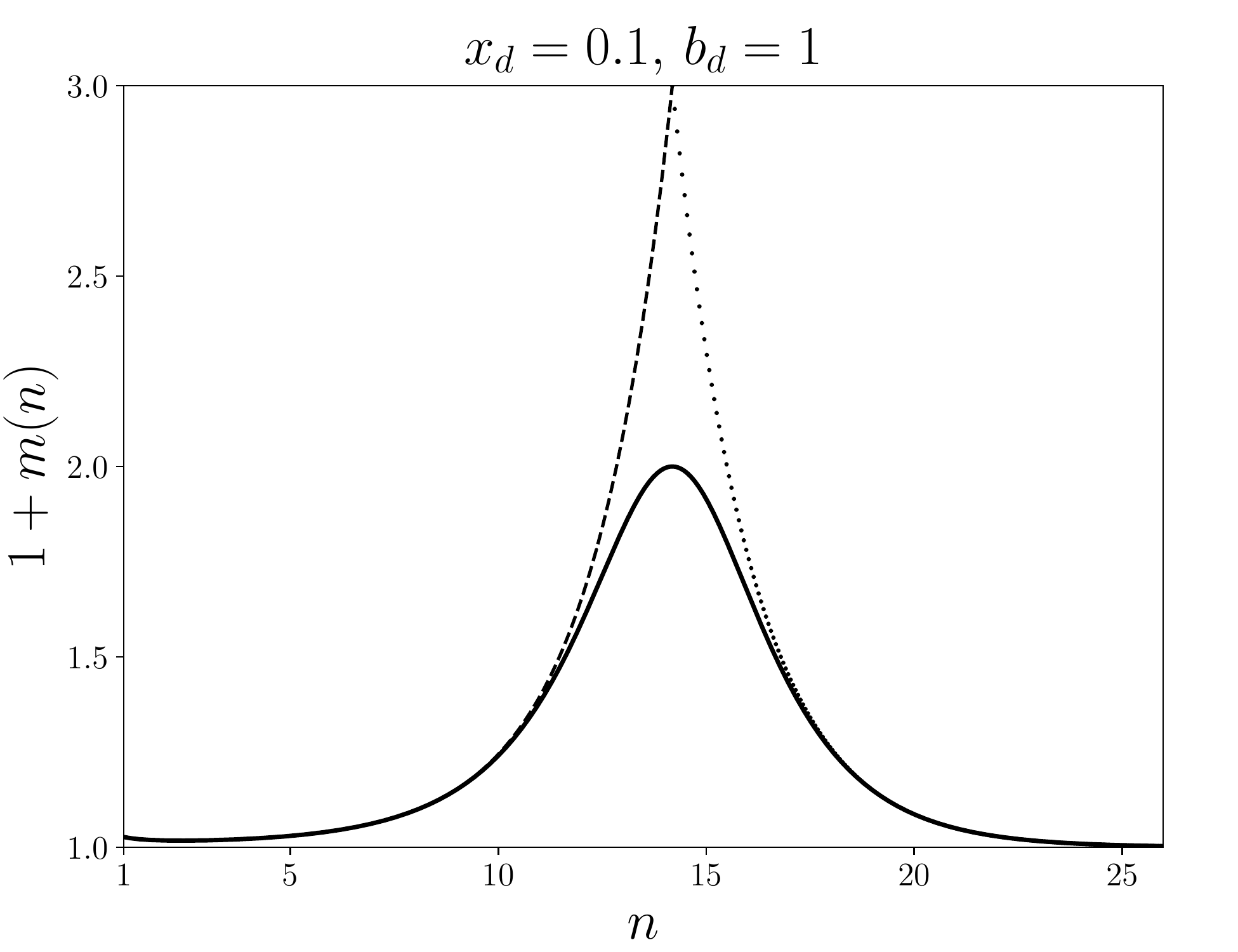}
\caption{Modulating function as a function of the number of oscillations $n$ of 
the survival probability.}\label{f11}
\end{figure}

Finally, the figure \ref{f11} plots $1+m(n)$, where $m(n)$ is given by Eq. \eqref{e7.8-13} 
and written in terms of $\eta(n)$ as $m(n) = 2/(\eta(n) + \eta(n)^{-1})$
where $m(n)$ (black line) has been computed using Eq. \eqref{e7.8-13} 
and using the approximations deduced for $\eta$. The dashed line is 
when $m(n)\approx 2/\eta$ in the exponential region and the dotted 
line is when $m(n)\approx 2\eta$ in the power law region.   
\section{Effect of other poles on $P_e(t)$}
\begin{figure}[hbt!]
\centering
\fbox{\includegraphics[scale=1.3]{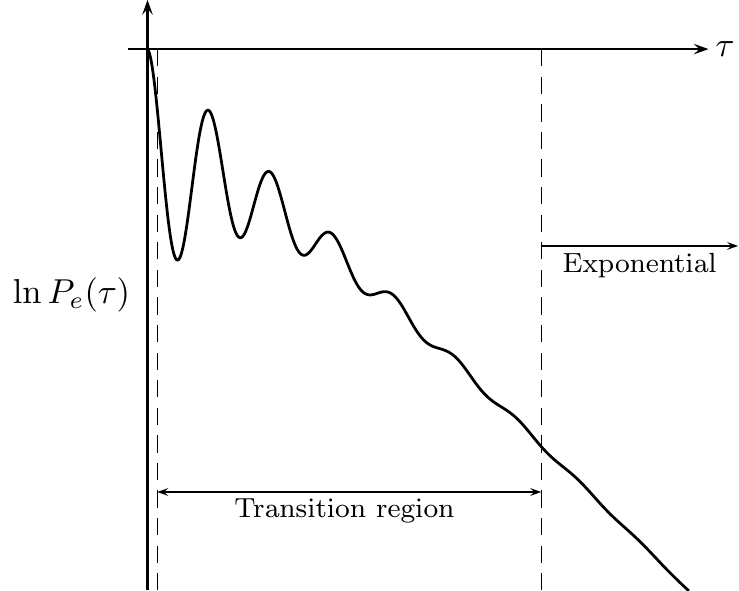}}
\caption{Sketch of the transition of $P_e(\tau)$ to the 
exponential law of the dominant pole.}\label{f3}
\end{figure}

An interesting and typical case is a system whose survival probability can be 
written approximately as a sum of the exponential terms only. 
After certain time, this sum should reduce to one term associated to the dominant isolated resonance. In order to see how this reduction law happens, we write Eq. \eqref{e7.3} such that the dominant pole is explicit, i.e.,
\begin{multline}\label{e7.9}
P_e(t)=|\bar{R}(z_d)|^2e^{-\om_dt}\EL{1+2\sum_{s>d}{\Re{\GC{\frac{\bar{R}(z_s)}{\bar{R}(z_d)}e^{-i(\sig_s-\sig_d)t}}}\exp{\GP{-\frac{\om_s-\om_d}{2}\,t}}}}\\
+\sum_{s'\neq d}{|\bar{R}(z_s')|^2e^{-\om_{s'}t}\EL{1+2\sum_{s>s'}{\Re{\GC{\frac{\bar{R}(z_s)}{\bar{R}(z_s')}e^{-i(\sig_s-\sig_s')t}}}\exp{\GP{-\frac{\om_s-\om_s'}{2}\,t}}}}},
\end{multline}  
and we can approximate the exponential survival amplitude by taking all the 
terms associated to the dominant pole:
\begin{equation}\label{e7.10}
P_e(t)\approx|\bar{R}(z_d)|^2e^{-\om_dt}\EL{1+2\sum_{s>d}{\Re{\GC{\frac{\bar{R}(z_s)}{\bar{R}(z_d)}e^{-i(\sig_s-\sig_d)t}}}\exp{\GP{-\frac{\om_s-\om_d}{2}\,t}}}}=|\bar{R}(z_d)|^2e^{-\om_dt}M(t),
\end{equation}
where $M(t)$ is defined as the function
\begin{equation}\label{e7.11}
M(t)={1+2\sum_{s>d}{\Re{\GC{\frac{\bar{R}(z_s)}{\bar{R}(z_d)}e^{-i(\sig_s-\sig_d)t}}}\exp{\GP{-\frac{\om_s-\om_d}{2}\,t}}}}.
\end{equation}
The modulating function $M(t)$ (which should not be confused 
with the modulating function defined by Eq. \eqref{e7.8-6}) contains 
information about how the reduction occurs. As is expected, this process does not depend on the dominant pole only but it also requires the other poles of the density of states for its description. If we use a 
``mean field approach'' in the sense of approximating the sum in 
Eq. \eqref{e7.11} by taking the nearest pole to the dominant one, 
the modulating function takes the form:
\begin{equation}\label{e7.12}
M(t)\approx{1+2{\Re{\GC{\frac{\bar{R}(z_{d+1})}{\bar{R}(z_d)}e^{-i(\sig_{d+1}-\sig_d)t}}}\exp{\GP{-\frac{\om_{d+1}-\om_d}{2}\,t}}}},
\end{equation}
or expressing the time in dimensionless (lifetime) units of the dominant 
resonance, i.e., $\tau=\om_dt$, we have, 
$P_e(\tau) = |\bar{R}(z_d)|^2e^{-\tau}M(\tau),\label{e7.13}$, where, 
\begin{equation}
M(\tau)=1+2\Re{\EC{\frac{\bar{R}(z_{d+1})}{\bar{R}(z_d)}\exp{\GP{-i\frac{\sig_{d+1}-\sig_d}{\om_d}\,\tau}}}}\exp{\EC{-\GP{\frac{\om_{d+1}}{\om_d}-1}\frac{\tau}{2}}}.\label{e7.18}
\end{equation}
From Eq. \eqref{e7.18}, we infer that the survival probability experiment an oscillation with a frequency $\om_t=(\sig_{d+1}-\sig_d)/\om_d$, and the survival amplitude goes to the exponential term associated to the dominant pole at a rate of  $2(\om_{d+1}/\om_d-1)^{-1}$ per lifetime unit 
(see Fig. \ref{f3}).

\end{document}